\documentclass[aps,pre,twocolumn,superscriptaddress,showpacs]{revtex4}
\usepackage{amsmath,amssymb,amsfonts}
\usepackage{bbm}
\usepackage{graphicx}
\newcommand\intd{\mathrm{d}}
\newcommand\ImgI{\mathrm{i}}

\newcommand\Exp{\mathrm{e}}

\newcommand\DET{\mathrm{det}}
\newcommand\TR{\mathrm{tr}}

\newcommand\ONE{\mathbbm{1}}
\newcommand\IM{\mathrm{Im}\,}
\newcommand\RE{\mathrm{Re}\,}
\begin{document}

\title{Universal spectral statistics in Wigner-Dyson, chiral and Andreev
  star graphs I: construction and numerical results}

\author{Sven Gnutzmann} 
\email{sven@gnutzmann.de}
\affiliation{Institut f\"ur Theoretische Physik, Freie Universit\"at
  Berlin, Arnimallee 14, 14195 Berlin, Germany} 
\author{Burkhard Seif}
\email{bseif@thp.uni-koeln.de} 
\affiliation{Institut f\"ur
  Theoretische Physik, Universit\"at zu K\"oln, Z\"ulpicher Str.\ 77,
  50937 K\"oln}
 
\begin{abstract} 
  
  In a series of two papers we investigate the universal spectral
  statistics of chaotic quantum systems in the ten known
  symmetry classes of quantum mechanics.
  In this first paper we  focus on the construction
  of appropriate ensembles of 
  star graphs in the ten symmetry classes. A
  generalization of the Bohigas-Giannoni-Schmit
  conjecture is given that covers all these
  symmetry classes. The conjecture
  is supported by numerical
  results that demonstrate
  the fidelity of the spectral statistics of
  star graphs to the corresponding
  Gaussian random-matrix theories. 

\end{abstract}      

\pacs{0.5.45.Mt,0.3.65.-w,74.50.+r}

\maketitle

\section{Introduction}

Based on earlier ideas of Wigner \cite{Wigner}
Dyson introduced a three-fold
classification of quantum systems according to their behavior under 
time-reversal and spin rotation \cite{Dyson}.  
This symmetry classification turned out to be very useful, for
instance 
in semiclassical, disordered and random-matrix
approaches to complex quantum systems. 
The success of random-matrix theory is based on universal features 
in spectra of complex quantum systems. While not capable of 
predicting single eigenvalues random-matrix theory has become one
of the key ingredients in predicting physical features that depend
on non-trivial spectral statistics \cite{Porter,Mehta}. 
In each symmetry class
various universality classes have been identified -- each described
by some ensemble of random matrices. Most prominent are the three
Gaussian ensembles GUE, GOE and GSE. They define the 
\emph{ergodic universality classes} and they have been applied 
successfully to a wide range of quantum systems (see the recent
review \cite{Guhr} for an overview and
further references). 

Recently the three-fold classification has been extended to
a ten-fold classification by incorporating spectral mirror symmetries.
This lead to seven novel symmetry classes 
\cite{chiral,Altland,Zirnbauer}. 
They are partly realized for a Dirac particle
in a random gauge field, and for
quasi-particles in disordered superconductors or
normalconducting-superconducting hybrid systems.
In the presence of a spectral mirror symmetry the spectrum is 
symmetric with respect to one point $E_0$: if
$E_0+E$ is in the spectrum so is $E_0-E$. The invention of the novel
classes has become necessary due to the impact such a symmetry
has on spectral correlations. These new universal features
appear near the symmetry point $E_0$ and they can be described
by defining random-matrix ensembles which incorporate the corresponding 
spectral mirror symmetry.

It has been
conjectured by Bohigas, Giannoni and Schmit that the spectra
of classically chaotic systems display the spectral fluctuations
described by the three Gaussian Wigner-Dyson ensembles
of random-matrix theory \cite{BGS}.
Though the fidelity to the universal predictions of
random-matrix theory have an overwhelming support by 
both experimental and numerical data the physical
basis of universality is not completely understood.

Quantum graphs have been introduced
by Kottos and Smilansky 
\cite{Kottos}
as simple quantum models with an exact semiclassical
trace formula for the density of states which is expressed as a sum
over periodic orbits on the graph.
They have since become an important
tool in the semiclassical approach to universality. 
In this series of papers we will construct star graphs
for all ten symmetry classes and investigate their spectral
statistics both numerically and analytically. 
While the following paper \cite{usII} is devoted to a semiclassical
periodic-orbit approach this paper focusses on the construction
of appropriate star graphs and some numerical results.

We start with 
giving a short introduction to the ten symmetry classes
in section \ref{sec:symmetry}
with all details needed for the subsequent construction 
of star graphs.
In the following section \ref{sec:universality}
on spectral statistics we introduce the spectral
form factors, review the results of Gaussian random-matrix theory
for the ten symmetry classes and generalize the Bohigas-Giannoni-Schmit
conjecture. 
After a general introduction to quantum graphs
in section \ref{sec:stargraphs} we construct one ensemble
of star graphs for each of the ten symmetry classes. 
Numerical results then show the fidelity of these 
ensembles to the predictions of the
Gaussian random-matrix ensembles.

\section{The ten symmetry classes of quantum systems}
\label{sec:symmetry}

In quantum mechanics most symmetries are described by some unitary
operators $\mathcal{U}$ that commute with the Hamilton operator
$\mathcal{H}=\mathcal{U} \mathcal{H} \mathcal{U}^\dagger$.  Thus the operators
$\mathcal{U}$ (or its hermitian generators) describe constants of
motion and they lead to a block diagonal form of the Hamilton matrix in an
eigenbasis of $\mathcal{U}$.   
If enough constants of motion
$\mathcal{U}_i$ are available such that the corresponding 
hermitian generators form a complete set of
\emph{commuting} observables the Hamilton operator is eventually
diagonalized in the common eigenbasis of the symmetry operators
(or their generators). 
However, for any hermitian Hamilton operator $\mathcal{H}$
there is always a complete set of commuting hermitian 
operators $\mathcal{P}_i$
which also commute with the Hamilton operator (e.g.~projectors
on eigenstates). In some sense the notion of
symmetry in this wide sense is obsolete for a single
quantum system. However, in most cases such a set of commuting operators
will not have any corresponding classical observable and will only
apply to a single system.

It is more appropriate for our purposes to consider a family 
or class of quantum systems.
Such a family may arrise 
by varying some physical parameters (like the
strength of an applied magnetic field) or, for disordered
systems, by an ensemble of random potentials. In the derivation
and application of semiclassical methods one formally considers
the asymptotics $\hbar\rightarrow 0$ which is
equivalent to a family of operators with fixed $\hbar$ but some varying 
physical parameters. 

A unitary operator $\mathcal{U}_i$ is a unitary symmetry
of a class of systems if it commutes with all Hamilton operators 
in that class. This notion avoids ``symmetry'' operators that would
rely on a diagonalization of the Hamilton operator. In semiclassics
a unitary symmetry will have a classical correspondence. In the
sequel we will 
consider each Hamilton operator as a member of a class
without explicitly referring to it.

If a family of systems has a unitary 
symmetry all its Hamilton operators
can be brought to a blockdiagonal form. Each block can be
regarded as a new Hamilton operator on a reduced Hilbert space.
Let us assume that the Hilbert space is completely reduced
such that there are no more unitary symmetries.
What types of symmetry may such a reduced
quantum system still have? What are the possible
structures of the Hamilton operator
(or the Hamilton matrix) and what are the consequences on its spectrum and
its eigenvectors?  Such questions were for the first time addressed 
and partially answered by
Wigner and Dyson \cite{Wigner,Dyson}.
Dyson proposed a symmetry classification based on the behavior
of quantum systems under time-reversal and spin rotation. 
This lead to three symmetry classes (the
three-fold way): \textit{i.} systems that are not time-reversal invariant,
\textit{ii.} 
time-reversal invariant spin-less particles, and \textit{iii.} time-reversal
invariant particles with spin $s=\frac{1}{2}$. Time-reversal symmetry has
immediate consequences on the form of the Hamilton operator:
spin-less particles can be described by real symmetric Hamilton
matrices in a time-reversal invariant basis, while systems without
time-reversal invariance do not have any canonical basis and
the Hamilton matrix remains complex.
The influence of the symmetry class on spectral properties such as
level repulsion has been investigated extensively within the field of
random matrix theory \cite{Guhr,Porter,Mehta}. 
We will give more details on random matrix theory
in section \ref{sec:universality}.

Recently the Wigner-Dyson symmetry classification has been extended to
a ten-fold way by including all different
types of mirror symmetries in the
spectrum \cite{chiral, Altland, Zirnbauer}. 
In the presence of a spectral mirror symmetry
every eigenvalue
$E_0+E$ has a partner eigenvalue at $E_0-E$ (we will set $E_0=0$ 
in the sequel without loss
of generality). 
Below we will describe the various ways a mirror
symmetry may arise and be combined with time-reversal invariance.
This
leads to the seven novel symmetry classes.  
As shown in \cite{Zirnbauer}
there is a one-to-one correspondence between Cartan's ten-fold classification
of Riemannian symmetric spaces and the ten symmetry classes of
quantum systems. We will use the
convention to adopt the names
given by Cartan to the different classes of symmetric spaces for the
according symmetry classes.

The novel symmetry classes are partly realized for Dirac fermions in
a random potential (the \emph{chiral classes}) \cite{chiral} and partly for
quasi-particles in mesoscopic superconductors or
superconducting-normalconducting (SN) hybrid systems (the
\emph{Andreev classes}). It is  
possible to construct much more general
systems in the appropriate symmetry classes, 
e.g.~two coupled spins or a generalized version of the Pauli
equation (which includes the Bogoliubov-de-Gennes equation as a
special case), and quantum graphs.  Quantum maps which incorporate
the corresponding symmetries have been discussed recently
\cite{henning}. Due to their simplicity graphs
will be the focus of this work.  The following discussion of
symmetry classes is summarized in table \ref{table1}.

\subsection{Time-reversal invariance}
\label{sec:time_rev_symm}

Quantum systems obey generalized time-reversal symmetry if there
is an anti-unitary operator $\mathcal{T}$ -- the \emph{generalized
  time reversal operator} -- that first, commutes with the Hamilton
operator
\begin{equation}
  \left[\mathcal{H},\mathcal{T}\right]=0
  \label{eq:timerevinv}
\end{equation}
and second \footnote{In general, $\mathcal{T}^2=\Exp^{\ImgI\phi}\ONE$
is the condition that any state is invariant upto a phase when it is 
time-reversed twice. The anti-unitarity of $\mathcal{T}$ then leads
to $\Exp^{\ImgI\phi}=\pm 1$.}, obeys
\begin{equation}
  \mathcal{T}^2=\pm \ONE.
  \label{eq:time_rev_phase}
\end{equation}  
Anti-unitarity implies 
\textit{i.}
anti-linearity $\mathcal{T} \left(\alpha
  |\xi\rangle +\beta |\nu\rangle \right)= \alpha^* |\mathcal{T} \xi
\rangle+ \beta^* |\mathcal{T} \nu \rangle$ and \textit{ii.} 
$\langle
\mathcal{T} \xi|\mathcal{T} \nu\rangle= \langle \xi|\nu\rangle^*$.

For time-reversal invariant systems $\mathcal{T}$ changes the
direction of time when applied to the Schr\"odinger equation. 
Equivalently,
when $\mathcal{T}$ is applied to the retarded Green's operator
\begin{equation}
  \mathcal{G}_{+}(E)=\frac{1}{E+\ImgI \epsilon-\mathcal{H}}
\end{equation}
one gets
\begin{equation}
  \mathcal{T}\mathcal{G}_{+}(E)\mathcal{T}^{-1}=
  \frac{1}{E-\ImgI \epsilon-\mathcal{H}}=\mathcal{G}_+(E)^\dagger\equiv
  \mathcal{G}_{-}(E),
  \label{eq:green_timerev}
\end{equation}
which is just the advanced Green's operator.

Time-reversal symmetry also effects other dynamic operators --
such as the unitary time evolution operator
\begin{equation}
  \mathcal{U}(t)=\Exp^{\ImgI \mathcal{H} t/\hbar}.
\end{equation} 
Scattering problems can often be described by some unitary
operator $\mathcal{S}(E)$ that connects incoming and outgoing
states of energy $E$.  Time-reversal invariance leads to
\begin{equation}
  \begin{split}
    \mathcal{T}\mathcal{U}(t)\mathcal{T}^{-1}&=\mathcal{U}(-t)=
    \mathcal{U}(t)^\dagger\\
    \mathcal{T}\mathcal{S}(E)\mathcal{T}^{-1}&=\mathcal{S}(E)^\dagger.
  \end{split}
  \label{eq:unitary_timerev}
\end{equation}
These equations also define time-reversal symmetry for quantum maps. 
The transformation of
the time development operator follows immediately from the condition
\eqref{eq:timerevinv} on the Hamiltonian.  In scattering problems
$\mathcal{S}(E)$ can be related to a unitary
combination of Green's functions -- for definiteness consider
$\mathcal{S}(E)=\mathcal{G}_+(E)\mathcal{G}_-(E)^{-1}=\ONE-
  2\,\epsilon\,\ImgI\,\mathcal{G}_+(E)$
and equation \eqref{eq:unitary_timerev} follows from the
transformation \eqref{eq:green_timerev} of $\mathcal{G}_+(E)$.

We have used the term \emph{generalized} time-reversal operator
because $\mathcal{T}$ need not be the well-known conventional 
time-reversal operator.  For a particle in $\mathbbm{R}^3$ the anti-unitary
conventional time-reversal operator obeys
\begin{alignat}{3}
  &\mathcal{T}_{\text{conv}}\, \vec{p}\,
  \mathcal{T}_{\text{conv}}^{-1}
  &=&& - \vec{p}\nonumber\\
  &\mathcal{T}_{\text{conv}}\, \vec{x}\,
  \mathcal{T}_{\text{conv}}^{-1}
  &=&& \vec{x}\\
  &\mathcal{T}_{\text{conv}}\, \vec{s}\,
  \mathcal{T}_{\text{conv}}^{-1} &=&& - \vec{s}
  \nonumber
\end{alignat}
where $\vec{s}$ is the particle spin.  
This conventional time-reversal operator obeys
$\mathcal{T}_{\mathrm{conv}}^2=\ONE$ if the spin quantum number is
integer $s=0,1,2,\dots$, and $\mathcal{T}_{\mathrm{conv}}^2=-\ONE$ if
the spin is half-integer $s=\frac{1}{2},\frac{3}{2},\dots$. Thus the
most relevant and simplest realizations are for spin-less
($\mathcal{T}_{\mathrm{conv}}^2=\ONE$) and spin-$\frac{1}{2}$
($\hat{T}_{\mathrm{conv}}^2=-\ONE$) particles. 

When a given quantum system is 
studied one should be aware that a generalized time-reversal operator
may still exist which commutes with the Hamiltonian while the
conventional time-reversal operator may not commute with
$\mathcal{H}$. 

The consideration of time-reversal symmetries leads to three symmetry
classes: either a system is not time-reversal invariant, or it is 
time-reversal invariant -- in the latter case the time-reversal operator
either obeys $\mathcal{T}^2=\ONE$ or $\mathcal{T}^2=-\ONE$.  These
classes have been called \emph{Wigner-Dyson classes} and their impact
on the form of Hamilton matrices and universal spectral features will
be discussed further in section \ref{sec:Wigner-Dyson} and
\ref{sec:universality}.  Additional spectral mirror symmetries lead to
the novel symmetry classes to be discussed below.

Kramers' degeneracy occurs
in time-reversal invariant quantum systems 
with $\mathcal{T}^2=-\ONE$. 
If $|\chi\rangle$ is an eigenvector with
eigenvalue $E$,
then due
to time reversal invariance
$|\mathcal{T}\chi\rangle\equiv\mathcal{T}|\chi\rangle$ is an
eigenvector with the same eigenvalue $E$. It is
straight forward to show that $|\mathcal{T}\chi\rangle$ is orthogonal
to $|\chi\rangle$ using the properties of the time reversal symmetry
operator. 
This degeneracy is well known for spin-$\frac{1}{2}$
particles with conventional time-reversal symmetry.

\subsection{Spectral mirror symmetries \label{sec:mirror_symm}}

A quantum system has a spectral mirror symmetry if the spectrum is
symmetric: for every eigenvalue $E>0$ there is another eigenvalue
$-E<0$. In general, there may be some vanishing eigenvalues $E=0$. 
We will now discuss the symmetry 
operators related to such spectral mirror symmetries.
 
According to a theorem by Wigner any symmetry operation on
Hilbert space is either represented by a unitary operator
$\mathcal{P}$ or an anti-unitary operator $\mathcal{C}$. Now take any
eigenstate $|\nu\rangle$ such that
$\mathcal{H}|\nu\rangle=E|\nu\rangle$ -- it is obvious that spectral
mirror symmetry demands that either
$\mathcal{H}\mathcal{P}|\nu\rangle=-E\mathcal{P}|\nu\rangle$ or
$\mathcal{H}\mathcal{C}|\nu\rangle=-E\mathcal{C}|\nu\rangle$.  This
condition on an eigenstate leads eventually to the condition that the
Hamilton operator anti-commutes with either a unitary or an
anti-unitary symmetry operator
\begin{align}
  \left[\mathcal{P},\mathcal{H}\right]_+&=0&\text{or}&&
  \left[\mathcal{C},\mathcal{H}\right]_+&=0.
  \label{eq:spectralmirror}
\end{align}  
As an additional condition any state should
be invariant upto a phase factor when acted upon
twice with $\mathcal{P}$ or $\mathcal{C}$. It can be shown that
it suffices to consider 
\begin{align}
  \mathcal{P}^2&=\pm \ONE&\text{or}&&
  \mathcal{C}^2=\pm \ONE.
\end{align} 
If a system is not time-reversal invariant one can
always choose $\mathcal{P}^2=\ONE$
while in presence of time
reversal symmetry either $\mathcal{P}^2=\ONE$ or
$\mathcal{P}^2=-\ONE$.

Spectral mirror symmetries relate Green's operators at energy
$E$ and $-E$
\begin{equation}
  \begin{split}
    \mathcal{P}\mathcal{G}_+(E)\mathcal{P}^{-1}&=-\mathcal{G}_-(-E)\\
    \mathcal{C}\mathcal{G}_+(E)\mathcal{C}^{-1}&=-\mathcal{G}_+(-E).
  \end{split}
\end{equation}
For scattering problems this leads to
\begin{equation}
  \begin{split}
    \mathcal{P}\mathcal{S}(E)\mathcal{P}^{-1}&=\mathcal{S}(-E)^\dagger\\
    \mathcal{C}\mathcal{S}(E)\mathcal{C}^{-1}&=\mathcal{S}(-E)
  \end{split}
  \label{eq:mirrorsym_scatt}
\end{equation}
and for the time development operator to
\begin{equation}
  \begin{split}
    \mathcal{P}\mathcal{U}(t)\mathcal{P}^{-1}&=\mathcal{U}(-t)
    =\mathcal{U}(t)^\dagger\\
    \mathcal{C}\mathcal{U}(t)\mathcal{C}^{-1}&=\mathcal{U}(t).
  \end{split}
  \label{eq:mirrorsym_timedev}
\end{equation}

The seven novel symmetry classes are obtained by all possible
combinations of a spectral mirror symmetry with time-reversal symmetry
(with the additional requirement that
$\left[\mathcal{P},\mathcal{T}\right]=0$ or
$\left[\mathcal{C},\mathcal{T}\right]=0$ if both symmetries are
supposed to hold). First, there are three novel
symmetry classes that are not time reversal invariant: either
there is a unitary operator with $\mathcal{P}^2=\ONE$ or an anti-unitary
with $\mathcal{C}^2=\pm \ONE$. 
In time-reversal invariant systems one has both unitary and anti-unitary
spectral mirror symmetry operators: if a unitary operator $\mathcal{P}$ 
anti-commutes with the Hamilton operator
so does the
anti-unitary operator $\mathcal{C}=\mathcal{P}\mathcal{T}$. 
As $\mathcal{T}^2=\pm\ONE$ and $\mathcal{C}^2=\pm\ONE$
this leads to
four symmetry classes that combine time-reversal symmetry with
spectral mirror symmetry: if $\mathcal{T}^2=\ONE$ either
$\mathcal{C}^2=\ONE$ ($\mathcal{P}^2=\ONE$) or $\mathcal{C}^2=-\ONE$
($\mathcal{P}^2=-\ONE$), if $\mathcal{T}^2=-\ONE$ either
$\mathcal{C}^2=-\ONE$ ($\mathcal{P}^2=\ONE$) or $\mathcal{C}^2=\ONE$
($\mathcal{P}^2=-\ONE)$.

For historic reasons these seven classes have been split into two
groups, the first group is given by the three \emph{chiral classes} -- the
ones that have a unitary mirror symmetry with $\mathcal{P}^2=\ONE$.
Their importance has first been observed in investigations of Dirac
fermions in a random gauge field where the spectral symmetry is related
to chirality.  For this reason we will call $\mathcal{P}$ a
\emph{chiral symmetry operator} though in general 
$\mathcal{P}$ need not be related to chirality.  The four
remaining classes have mainly been discussed in connection to
mesoscopic disordered superconductors or superconducting-normalconducting
hybrid systems where the
anti-unitary mirror symmetry is connected to electron-hole
conjugation. For this reason we call $\mathcal{C}$ a \emph{charge
  conjugation symmetry operator}, though again, in general
$\mathcal{C}$ need not be related
to charge conjugation at all.  Since Andreev reflection is a main
ingredient in the dynamics of  superconducting-normalconducting
hybrid systems we will call these
classes \emph{Andreev classes}.  The detailed discussion of these
symmetry classes and their impact on universal spectral features will
be discussed in sections \ref{sec:chiral}, \ref{sec:Andreev} and
\ref{sec:universality}.
\begin{table}
  \begin{center}
    \begin{tabular}{ccccc}
      \hline
      symmetry&$\qquad$&$\qquad$&$\qquad$&symmetric\\
      class &
      $\mathcal{T}$& $\mathcal{P}$ & $\mathcal{C}$ & space
      \\
      \hline
      \hline
      $A$&0&0&0& $U(N)$\\
      $A$I&+1&0&0& $U(N)/O(N)$\\
      $A$II&-1&0&0&$U(2N)/Sp(N)$\\
      $A$III&0&+1&0&$U(p+q)/U(p)\times U(q)$\\
      $BD$I&+1&+1&+1&$SO(p+q)/SO(p)\times SO(q)$\\
      $C$II&-1&+1&-1&$Sp(p+q)/Sp(p)\times Sp(q)$\\
      $C$&0&0&-1&$Sp(N)$\\
      $C$I&+1&-1&-1&$Sp(N)/U(N)$\\
      $BD$ ($D$) &0&0&+1&$SO(N)$\\
      $D$III &-1&-1&+1&$SO(2N)/U(N)$\\
      \hline
    \end{tabular}
  \end{center}
  
  \caption{   \label{table1}
    The ten symmetry classes of quantum systems. If a symmetry class
    obeys time-reversal symmetry or a spectral mirror symmetry
    the entry $\pm 1$ in the corresponding column indicates
    if the symmetry operator squares to $\pm \ONE$. 
    The entry $0$ indicates that the corresponding symmetry is broken.
    The last column gives the corresponding Riemannian symmetric space 
    (of compact type).
  }
\end{table}

\subsection{\label{sec:canonicalbases} Explicit form of scattering 
  matrices for each symmetry class}

Time-reversal and spectral mirror symmetries 
restrict the form of Hamilton and scattering
matrices due
to the relations \eqref{eq:timerevinv}, \eqref{eq:unitary_timerev},
\eqref{eq:spectralmirror} and \eqref{eq:mirrorsym_scatt}. By choosing
an appropriate Hilbert space basis for each symmetry class the
symmetry operators are represented by a simple matrix 
(combined with the complex conjugation operator 
for anti-unitary operators). These determine the explicit 
form of scattering matrices for each symmetry class. 

Note, that the following derivation of the scattering matrices
depends on the choice of the basis. There are many choices
for the Hilbert space basis in which the symmetry operators
have a simple form. As a consequence many of
the following identities are only valid in that special basis.
Especially the  ``complex conjugation operator'' 
$\mathcal{K}$ is defined
with respect to a given basis. However, one may show that the
bases chosen here can always
be constructed from the general properties of the time-reversal
and spectral mirror symmetries.
Our choice of basis is biased by their later 
application to star graphs in section \ref{sec:stargraphs}.

In addition, some symmetry classes have a 
further division into subclasses.
Though we will mention all subclasses 
we will only give the scattering matrix in one of the
subclasses.

\subsubsection{The Wigner-Dyson classes\label{sec:Wigner-Dyson}}

Quantum systems without spectral mirror symmetries belong to one of
the three Wigner-Dyson classes $A$, $A$I or $A$II.

Class $A$ contains quantum systems that are not time-reversal
invariant.  There is no preferred basis in Hilbert space and the
scattering matrix $\mathcal{S}(E)$ may be any 
unitary $N \times N$ matrix.

A time-reversal invariant system belongs either to class
$A$I if $\mathcal{T}^2=\ONE$ or to class $A$II if
$\mathcal{T}^2=-\ONE$.

In class $A$I there are
time-reversal invariant bases such that
$\mathcal{T}|i\rangle=|i\rangle$ for any basis state.
In any such basis the time-reversal symmetry operator is 
represented by the complex conjugation
operator
\begin{equation}
  \text{$A$I}: \qquad \mathcal{T} \equiv \mathcal{K},
\end{equation} 
where the complex conjugation operator acts on a general state by
complex conjugation of the coefficients
$\mathcal{T} \sum_{i=1}^N a_i |i\rangle \equiv \mathcal{K} 
\sum_{i=1}^N a_i  |i\rangle = \sum_{i=1}^N a_i^* |i\rangle$. 

The condition \eqref{eq:unitary_timerev} implies that a
scattering matrix is represented by a unitary symmetric
$N\times N$ matrix 
\begin{equation}
  \begin{array}{lr}
    \text{$A$I}: & \mathcal{S}(E)=\mathcal{S}(E)^\dagger.
  \end{array}
\end{equation}

For class $A$II there is no time-reversal invariant basis.
Instead, there are always bases in which
the time-reversal symmetry operator is be represented by
\begin{equation}
  \text{$A$II}:\qquad \mathcal{T} \equiv \mathcal{K}
  \begin{pmatrix}
    0 & - \ONE \\
    \ONE & 0
  \end{pmatrix}
  \label{eq:canon_timerevII}
\end{equation}
where $\ONE$ is the $N\times N$ identity matrix. 
Hilbert space has even
dimension due to Kramers' degeneracy.
In such a basis the scattering matrix has the form
\begin{equation}
  \begin{array}{lr}
    \text{$A$II}: & \mathcal{S}(E)=
    \begin{pmatrix}
      \scriptstyle
      \mathcal{X}_1(E) & \scriptstyle \mathcal{X}_2(E)\\
      \scriptstyle \mathcal{X}_3(E) & \scriptstyle \mathcal{X}_1(E)^T
    \end{pmatrix}
  \end{array}
  \label{eq:canon_S_AII}
\end{equation}
with complex $N \times N$ matrices $\mathcal{X}_i$ that satisfy
$\mathcal{X}_2(E)=-\mathcal{X}_2(E)^T$ and
$\mathcal{X}_3(E)=-\mathcal{X}_3(E)^T$ and are further restricted
by unitarity of $\mathcal{S}(E)$.

\subsubsection{The chiral classes}
\label{sec:chiral}

A system with a spectral mirror symmetry connected to a unitary
chiral symmetry operator $\mathcal{P}$ (with $\mathcal{P}^2=\ONE$)
falls into one of the three chiral symmetry classes $A$III, $BD$I or
$C$II.  

Since $\mathcal{P}^2=\ONE$ the eigenvalues are either $+1$ or
$-1$. In general, there will be $p$ positive and $q$ negative
eigenvalues. 
The number $\nu=|p-q|=0,1,2,\dots$ distinguishes
between different subclasses in each of the chiral classes
($\nu$ is always even
for class $C$II). The integer $\nu$ has impact on both
the form of Hamilton (or scattering) matrices and on
the spectral statistics.
Because $\mathcal{P}$ relates states with positive
energy to states with negative energy
there are $\nu$ vanishing energy eigenvalues due
to the chiral symmetry. 

We will focus on the subclasses with $\nu=0$ 
and set 
$p=q\equiv N$ in classes $A$III and $BD$I, $p=q=2N$ in class $C$II.
Hilbert space has even
dimension in all three classes.  
There are many bases that can be used as reference basis,
for example the one, where $\mathcal{P}$ is diagonal. Here, 
biased by our following construction of star graphs we choose
\begin{equation}
  \text{$A$III, $BD$I, $C$II}:\qquad
  \mathcal{P}=
  \begin{pmatrix}
    0 & \ONE\\
    \ONE & 0
  \end{pmatrix}
 \label{eq:canon_chiral}
\end{equation} 
which can be obtained from the diagonal representation by a simple
rotation. 

The chiral class $A$III contains systems without
additional time-reversal invariance. The other two
chiral symmetry classes are time-reversal invariant
with $\mathcal{T}^2=\ONE$ for class $BD$I and
$\mathcal{T}^2=-\ONE$ for class $C$II.
In class $BD$I one
may always choose a time-reversal symmetry operator
of the form
\begin{equation}
  \text{$BD$I}:\qquad \mathcal{T}\equiv \mathcal{K}
\end{equation}
which commutes with the chiral symmetry operator $\mathcal{P}$
\eqref{eq:canon_chiral}.
In class $C$II one may choose 
\begin{equation}
  \text{$C$II}:\qquad \mathcal{T}=\mathcal{K}
  \begin{pmatrix}
    0 &-\ONE&0&0\\
    \ONE&0&0&0\\
    0&0&0&-\ONE\\
    0&0&\ONE&0
  \end{pmatrix}
\end{equation}
which also commutes $\mathcal{P}$.

Due to the condition  \eqref{eq:mirrorsym_scatt}
a scattering matrix $\mathcal{S}(E)$
in class $A$III has the form
\begin{equation} 
  \begin{array}{lr}
    \text{$A$III}: & \mathcal{S}(E)=
    \begin{pmatrix}
      \scriptstyle
      \mathcal{X}_1(E) &  \scriptstyle \mathcal{X}_2(E)\\
      \scriptstyle \mathcal{X}_3(E) &  \scriptstyle \mathcal{X}_1(-E)^\dagger
    \end{pmatrix}
  \end{array}
  \label{eq:canon_S_AIII}
\end{equation}
where (besides unitarity) the $N \times N$ matrices 
$\mathcal{X}_i$ are further restricted
by $\mathcal{X}_2(E)=\mathcal{X}_2(-E)^\dagger$ and 
$\mathcal{X}_3(E)=\mathcal{X}_3(-E)^\dagger$.

In class $BD$I, due to 
time-reversal invariance \eqref{eq:unitary_timerev}, 
$\mathcal{S}(E)$ is symmetric, thus
\begin{equation} 
  \begin{array}{lr}
    \text{$BD$I}: & \mathcal{S}(E)=
    \begin{pmatrix}
      \scriptstyle\mathcal{X}_1(E) &  \scriptstyle\mathcal{X}_2(E)\\
      \scriptstyle\mathcal{X}_2(E)^T &  \scriptstyle \mathcal{X}_1(-E)^*
    \end{pmatrix}
  \end{array}
  \label{eq:canon_S_BDI}
\end{equation}
where $\mathcal{X}_1(E)=\mathcal{X}_1(E)^T$ and  
$\mathcal{X}_2(E)=\mathcal{X}_2(-E)^\dagger$. 

Finally,
in class $C$II, $\mathcal{S}(E)$ is a $4N \times 4N$ matrix 
of the form
\begin{equation}
  \begin{array}{lr}
    \!\text{$C$II}\!:\,& \!\mathcal{S}(E)\!=\!\!\!
    \begin{pmatrix}
      \!\scriptstyle\mathcal{X}_1(E)\! &  
      \!\scriptstyle\mathcal{X}_2(E)\!&
      \!\scriptstyle\mathcal{X}_3(E)\! &
      \!\scriptstyle\mathcal{X}_4(E)\!\!\!\\
      \!\scriptstyle\mathcal{X}_5(E)\!&
      \!\scriptstyle\mathcal{X}_1(E)^T\! &
      \!\scriptstyle\mathcal{X}_4(-E)^\dagger\!&
      \!\scriptstyle\mathcal{X}_6(E)\!\!\!\\
      \!\scriptstyle\mathcal{X}_6(E)^T\!&
      \!\scriptstyle -\mathcal{X}_4(E)^T\!&
      \!\scriptstyle\mathcal{X}_1(-E)^\dagger\!&
      \!\scriptstyle\mathcal{X}_5(-E)^\dagger\!\!\!\\
      \!\scriptstyle -\mathcal{X}_4(-E)^*\!&
      \!\scriptstyle\mathcal{X}_3(E)^T\!&
      \!\scriptstyle\mathcal{X}_2(-E)^\dagger\!&
      \!\scriptstyle\mathcal{X}_1(-E)^*\!\!\!
    \end{pmatrix}
  \end{array}
  \label{eq:canon_S_CII}
\end{equation}
with additional constraints $\mathcal{X}_2(E)=-\mathcal{X}_2(E)^T$,
$\mathcal{X}_3(E)=\mathcal{X}_3(-E)^\dagger$, 
$\mathcal{X}_5(E)=-\mathcal{X}_5(E)^T$
and $\mathcal{X}_6(E)=\mathcal{X}_6(-E)^\dagger$.

\subsubsection{The Andreev classes}
\label{sec:Andreev}

A quantum system with a spectral mirror symmetry that does
not belong to any of the chiral symmetry classes
belongs to one of the four Andreev classes $C$,
$C$I, $BD$ or $D$III. The spectral mirror symmetry for these
classes is related to an anti-unitary charge conjugation
operator $\mathcal{C}$ with $\mathcal{C}^2=-\ONE$
for $C$ and $C$I while $\mathcal{C}^2=\ONE$
for $BD$ and $D$III. The classes $C$ and $BD$ are not time-reversal
invariant while $C$I and $D$III are time-reversal invariant
with $\mathcal{T}^2=\ONE$ in $C$I and $\mathcal{T}^2=-\ONE$
in $D$III.

The classes $C$ and $C$I do not split into subclasses.
In appropriate $2N$-dimensional bases the charge conjugation 
operator can be represented as
\begin{equation}
  \text{$C$, $C$I}:\qquad \mathcal{C}= \mathcal{K}
  \begin{pmatrix}
    0 & -\ONE\\
    \ONE & 0
  \end{pmatrix}
  \label{eq:canon_chargeC}
\end{equation}
In contrast the classes $BD$ and $D$III fall into two
subclasses each. The symmetry class $BD$ allows for
either an even or odd dimensional Hilbert space. 
Due to spectral mirror symmetry there is always an eigenvalue on the
symmetry point $E=0$ in an odd-dimensional Hilbert space.
The subclass with odd (even) dimensional Hilbert space may be called
$BD$-odd(even). In the following we will restrict
ourselves to the even-dimensional case and
will follow the convention to call it symmetry class $D$.
Similarly $D$III falls into the two subclasses $D$III-odd
and $D$III-even. The dimension of the corresponding Hilbert spaces
is twice an odd or twice an even number. Spectral mirror symmetry
combined  with Kramers' degeneracy implies two
eigenvalues $E=0$ on the spectral symmetry point in class $D$III-odd.
In the sequel we will restrict to $D$III-even which is
physically more relevant.

An appropriate choice of basis in
the Hilbert space takes the charge conjugation operator 
of the symmetry classes $D$
and $D$III (we will not mention the ``even'' further) to
the form
\begin{equation}
  \begin{array}{lr}
    \text{$D$,$D$III}:& \mathcal{C}=\mathcal{K}
    \begin{pmatrix}
      0 & \ONE\\
      \ONE & 0
    \end{pmatrix},
  \end{array}
\end{equation} 
where $\ONE$ is the $N \times N$ identity matrix for class $D$
and the $2N \times 2N$ identity for class $D$III.

The time-reversal symmetry operators in the classes $C$I
and $D$III have the representations
\begin{equation}
  \begin{array}{lr}
     \text{$C$I}:&
     \mathcal{T}=\mathcal{K}
   \end{array}
\end{equation}
and
\begin{equation}
  \begin{array}{lr}
    \text{$D$III}: &
    \mathcal{T}=\mathcal{K}
    \begin{pmatrix}
      0 & -\ONE & 0 & 0\\
      \ONE &0 &0 &0\\
      0 & 0 & 0 & -\ONE \\
      0 & 0 & \ONE & 0
    \end{pmatrix}
  \end{array}
\end{equation}
in an appropriate basis -- here $\ONE$ is the $N \times N$
identity matrix in both equations. These representations 
commute with the corresponding representations of the charge
conjugation operators.

The conditions \eqref{eq:unitary_timerev} and 
\eqref{eq:mirrorsym_scatt}
to scattering matrices $\mathcal{S}(E)$ 
of the form 
\begin{equation} 
  \begin{array}{lr}
    \text{$C$, $C$I}: & \mathcal{S}(E)=
    \begin{pmatrix}
      \scriptstyle \mathcal{X}_1(E) & \scriptstyle \mathcal{X}_2(E)\\
      \scriptstyle -\mathcal{X}_2(-E)^* &\scriptstyle  \mathcal{X}_1(-E)^*
    \end{pmatrix}
  \end{array}
  \label{eq:canon_S_C}
\end{equation}
in classes $C$ and $C$I. 
There are no further restrictions 
on the complex $N \times N$ matrices $\mathcal{X}_i$
for class $C$ (apart from
unitarity).
Time-reversal invariance in class $C$I requires
$\mathcal{S}(E)$ to be symmetric, thus
$\mathcal{X}_1(E)=\mathcal{X}_1(E)^T$ and
$\mathcal{X}_2(E)=-\mathcal{X}_2(-E)^\dagger$.  

In the symmetry class $D$ the scattering matrix has the form
\begin{equation} 
  \begin{array}{lr}
    \text{$D$}: & \mathcal{S}(E)=
    \begin{pmatrix}
      \scriptstyle \mathcal{X}_1(E) &\scriptstyle  \mathcal{X}_2(E)\\
      \scriptstyle \mathcal{X}_2(-E)^* &\scriptstyle  \mathcal{X}_1(-E)^*
    \end{pmatrix}
  \end{array}
  \label{eq:canon_S_D}
\end{equation}
without further restrictions on the $N\times N$ matrices $\mathcal{X}_i$.

For class $D$III $\mathcal{S}(E)$ is a complex $4N\times 4N$ matrix of
the form
\begin{equation}
  \begin{array}{lr}
    \text{$D$III}\!:&  
    \mathcal{S}(E)\!\!=\!\!\!        
    \begin{pmatrix}
      \!\!\scriptstyle \mathcal{X}_1(E) \!& 
      \!\scriptstyle \mathcal{X}_2(E)\!& 
      \!\scriptstyle  \mathcal{X}_3(E) \!&
      \!\scriptstyle \mathcal{X}_4(E)\!\!\\
      \!\!\scriptstyle \mathcal{X}_5(E)\!& 
      \!\scriptstyle \mathcal{X}_1(E)^T \!& 
      \!\scriptstyle  \mathcal{X}_6(E)\!&
      \!\scriptstyle  \mathcal{X}_3(-E)^\dagger\!\!\\
      \!\!\scriptstyle  \mathcal{X}_3(-E)^*\!& 
      \!\scriptstyle \mathcal{X}_4(-E)^*\!& 
      \!\scriptstyle  \mathcal{X}_1(-E)^*\!&
      \!\scriptstyle  \mathcal{X}_2(-E)^*\!\!\\
      \!\!\scriptstyle  \mathcal{X}_6(-E)^*\!&
      \!\scriptstyle   \mathcal{X}_3(-E)^T\!&
      \!\scriptstyle   \mathcal{X}_5(-E)^*\!&
      \!\scriptstyle  \mathcal{X}_1(-E)^\dagger\!\!
  \end{pmatrix}
  \end{array}
  \label{eq:canon_S_DIII}
\end{equation}
with $\mathcal{X}_2(E)=-\mathcal{X}_2(E)^T$,
$\mathcal{X}_4(E)=-\mathcal{X}_4(-E)^\dagger$,
$\mathcal{X}_5(E)=-\mathcal{X}_5(E)^T$ and
$\mathcal{X}_7(E)=-\mathcal{X}_7(-E)^\dagger$.

\section{Universal spectral statistics}
\label{sec:universality}

In the previous chapter we have summarized the symmetry classification
of quantum systems. It is completely
general. We have not yet related it to universal spectral properties.
This will be done in this section.
In each symmetry class there are several universal regimes
with respect to their spectral statistics.
A universality class is a subset of a symmetry class which share
the same spectral statistics (or at least some 
universal spectral correlation
functions). The spectral statistics of a given universality class can
be described (and defined) by some ensemble of random matrices
(usually there will be a lot of different ensembles that share
the same universal spectral statistics).
In this paper we will focus on the \emph{ergodic}
universality classes that can be described by Gaussian ensembles 
of Hermitian matrices in each of the ten symmetry classes.
Note, that three chiral symmetry classes and the 
symmetry classes $BD$ and $D$III
fall into various subclasses -- as the universal
spectral statistics is different in each of these subclasses
they define different ergodic universality classes
in the same symmetry class.
As in the previous section we will only discuss one subclass
in each of these cases. In the symmetry classes
$A$, $A$I, $A$II, $C$ and $C$I there is \emph{one} unique ergodic
universality class. In the chiral classes $A$III, $BD$I and $C$II
we restrict to $|p-q|=0$ (see section \ref{sec:chiral}). Finally,
the classes $BD$ and $D$III have two subclasses 
(see section \ref{sec:Andreev}) and we will restrict ourselves to 
the subclasses $D$ ($BD$-even) and $D$III-even.

Let us mention that apart from 
the ergodic universality classes there are a lot of other
physically relevant universality classes within each symmetry class.
In random-matrix theory these correspond to ensembles 
which are not equivalent
to the Gaussian ensembles. For instance ensembles of banded or sparse
Hermitian matrices can describe quantum systems in a localized regime
\cite{Fyodorov}. 

In Andreev systems more specialized
random-matrix ensembles can describe the so-called hard gap 
in the quasi-particle excitation spectrum that appears when a
small part of the boundary of a
normalconducting chaotic billiard is coupled to a 
superconductor \cite{hardgap}. If no magnetic field is applied
the resulting combined electron-hole dynamics 
near the Fermi level is no longer
chaotic and the system does not belong to an ergodic universality
class.

\subsection{\label{sec:fluctuatingdos}
  The fluctuating part of the density of states}

To reveal universality in the statistics of quantum spectra the
system dependent mean density of states has
to be separated. This is done by writing the density of states as
a sum
\begin{equation}
  d(E)=\sum_i \delta(E-E_i)=d_{\mathrm{Weyl}}(E) +\delta d(E).
  \label{eq:dos}
\end{equation}
In presence of Kramers' degeneracy (symmetry classes
$A$II, $C$II, and $D$III) we define the density of states such that
every doubly degenerate energy is counted only once in the sum
$d(E)=\sum_i \delta(E-E_i)$. Let us also introduce a degeneracy factor
$g$, where $g=2$ for systems with Kramers' degeneracy and else $g=1$.

In equation \eqref{eq:dos} the first part $d_{\mathrm{Weyl}}(E)$ is
the average density of states which may be obtained by
counting all states $E_i$ in an interval
$E-\frac{E_I}{2}<E_i<E+\frac{E_I}{2}$, then the number $N_I$ of states
in that interval divided by $E_I$ is the average density of states
\begin{equation}
  d_{\mathrm{Weyl}}(E)= \frac{N_I}{E_I}.
\end{equation}
For this to be well-defined it is necessary to
choose $E_I$ 
self-consistently in range such that \textit{i.}
$N_I \gg 1$ which is
equivalent to taking the energy interval much larger the the mean
spacing $E_I \gg \Delta E=\frac{1}{d_{\text{Weyl}}}$, and \textit{ii.}
$E_I$ is small compared to compared to the scale on which
the resulting $d_{\text{Weyl}}$ changes.

In systems that allow for a classical limit
and one may consider the semiclassical regime. 
The scale $E_I$ is then chosen classically small 
($E_I\rightarrow 0$ as $\hbar \rightarrow 0$) but large compared to
the mean level spacing.
Thus the average density
of states is well-defined in the semiclassical regime.
It is given by Weyl's law
\begin{equation}
  d_{\text{Weyl}}(E)= \int\frac{\intd^f\mathbf{p}\intd^f\mathbf{q}}{
    \left( 2 g \pi\hbar \right)^f} 
  \delta(E-H_{\text{class}}(\mathbf{p},\mathbf{q}))
  \label{eq:weyl}
\end{equation}
where $H_{\text{class}}(\mathbf{p},\mathbf{q})$ is the classical
Hamilton function, and $f$ the number of freedoms. This equation shows
that the average density of states defined by Weyl's law is 
system dependent and universal features can only arise due to the
fluctuating part $\delta d(E)$. 
Note, that Weyl's law gives the mean
density of states on scales much larger than the mean level spacing.
In the presence of mirror symmetries the fluctuating part
$\delta d(E)$ may contribute 
to \emph{universal} features in the density
of states on the scale of the mean level spacing.

For classically chaotic (hyperbolic) systems the fluctuating part of
the density of states is given by Gutzwiller's trace formula
\cite{Gutzwiller} as a sum over periodic orbits of the classical
system
\begin{equation}
  \delta d (E)= \sum_{\mathrm{p.o.}\, \alpha} \frac{t_{\alpha}}{g \hbar \pi} 
  A_{\alpha}
  \cos{\frac{W_\alpha}{\hbar}}.
  \label{eq:Gutzwiller}
\end{equation}
Here, $t_\alpha$ is the primitive period of the orbit (the time needed
for a single traversal),
$A_\alpha=\frac{\Exp^{-\ImgI\mu_\alpha\frac{\pi}{2}}}{ \sqrt{|\DET
    M^{\mathrm{red}}_\alpha-\ONE}|}$ is the stability amplitude of the
periodic orbit ($M^{\text{red}}_\alpha$ is the \emph{reduced monodromy
  matrix} and $\mu_\alpha$ the \emph{Maslov index}) and
$W_\alpha=\oint_\alpha \mathbf{p}\, \intd \mathbf{q}$ is the (reduced)
action.  Note, that hyperbolic chaos is a strong condition on a
classical system -- all periodic orbits are
hyperbolically unstable and isolated in the energy shell. 

The energy scale for universal features is given by the mean level
spacing $\Delta E= \frac{1}{d_{\text{Weyl}}}$. Introducing rescaled
energies  $E=\epsilon \Delta E$ one obtains a density of
states 
\begin{equation}
  d(\epsilon)=1 +\delta d(\epsilon)
  \label{eq:unfolded_dos}
\end{equation}
for the \emph{unfolded} spectrum.

\subsection{\label{sec:gaussian_RMT}
  Gaussian ensembles of random-matrix theory}

Each 
ergodic universality class can be associated to a Gaussian ensemble
of random matrices.  
Within one class the Gaussian ensembles differ only by the
dimension of their matrices. The universal features of spectral
statistics are extracted in the limit of large matrices.

In each Gaussian ensemble the probability for a Hamiltonian matrix
$\mathcal{H}$ (with symmetries according to one of the ten symmetry
classes) has the form
\begin{equation}
  P(\mathcal{H}) \intd \mu(\mathcal{H}) = 
  \frac{1}{N}\,\Exp^{-A\,\TR \mathcal{H}^2}
  \intd \mu(\mathcal{H})
\end{equation} 
where $N$ is a normalization constant, $A$ is an overall scale that
fixes the mean level spacing, and the measure $\intd\mu(\mathcal{H})$
is given by $\prod \intd\RE\mathcal{H}_{ij} \,\intd \IM
\mathcal{H}_{ij}$ where the product runs over all independent elements
of $\mathcal{H}$.  

In general, on may denote the Gaussian ensemble for
the symmetry class $X$ by $X$-GE. We will use this notion for the 
Andreev classes.
Note, that for some symmetry classes one should distinguish
various ergodic universality classes.
As we have restricted our investigations to just one relevant subclass
we will use the name of the whole symmetry class for the Gaussian
ensembles. 

\subsection{\label{sec:formfactors} Spectral form factors}

Let us now define the statistical functions
that are in the center of our investigation.

For a physical system the following averages are either performed
over some system
parameters or over different parts of the spectrum.
We will always use unfolded spectra with unit mean
level spacing. Spectral
averaging is only possible if the universal results are
invariant under shifts of the energy $E \rightarrow E+E'$.  

There is an important difference between the
Wigner-Dyson classes where the universality was conjectured 
for a  single spectrum of one system and the remaining 
seven symmetry classes
where some universal features near energy $E=0$ can only be obtained
by averaging over different spectra. In the classical limit
one naturally obtains many spectra for the same physical system
by formally changing $\hbar$. Thus even for the seven 
novel symmetry classes
one may average over different spectra 
for the \emph{same physical system}.

We will be interested in the two simplest correlation functions and
their Fourier transforms. We will call the latter 
\emph{form factors}. The first correlation function is simply the 
averaged
fluctuating part of the density of states $\langle \delta d(\epsilon)
\rangle$.  If the spectral statistics is
invariant under shifts this expectation value must vanish (if
not it would be a constant over scales much larger than the mean level
spacing -- in contradiction to its definition). 
The spectral statistics near a spectral mirror
symmetry is not
invariant under energy-shifts. Non-trivial 
contributions to the mean fluctuating part of the density of states
may then arise. These have to appear 
on the scale of mean level spacing (else it would
be inconsistent with the separation of 
the density of states in $d_{\text{Weyl}}+\delta d$).

The
Fourier transform of the averaged fluctuating part of the density of
states is the \emph{first-order form factor}
\begin{equation}
  K_1(\tau)=2 \int_{-\infty}^{\infty} \intd \epsilon \;
  \Exp^{-\ImgI 2 \pi \epsilon \tau} \left\langle \delta d(\epsilon)
    \right\rangle
    \label{eq:K1}
  \end{equation}
where $t_H=\frac{2\pi \hbar}{\Delta E}$ is the Heisenberg time.
Inverting the Fourier transform one may represent the deviations from
Weyl's law in the expectation value for the density of states as
\begin{equation}
  \left\langle \delta d(\epsilon)
    \right\rangle=\int_0^\infty \intd \tau 
    \cos\left(2\pi\epsilon\tau\right)K_1(\tau).
\end{equation}
Note, that for $\tau>0$ the first-order form factor is the trace of
the time evolution operator $K_1(\tau)=\TR\, 
\Exp^{\ImgI \frac{H \tau t_H}{\hbar}}$.

The second-order correlation function is defined by
\begin{equation}
  C(\epsilon,\epsilon_0)=\left\langle \delta d(\epsilon_0+\epsilon/2) 
  \delta d(\epsilon_0-\epsilon/2) \right\rangle.
\end{equation}
If the spectral statistics is invariant under energy shifts it
only depends on the energy difference
$\epsilon$ -- averaging over different parts of the spectrum for a
given system is an average over $\epsilon_0$.  Its
Fourier transform with respect to
$\epsilon$ is the \emph{second-order form factor} 
\begin{equation}
    K_2(\tau)=\int_{-\infty}^\infty \intd \epsilon\, \Exp^{-\ImgI 2
      \pi \epsilon \tau} C(\epsilon,\epsilon_0)
  \label{eq:K2}
\end{equation}
where we have suppressed the possible dependency on $\epsilon_0$.

For physical spectra a time average
over a small time interval $\Delta \tau\ll 1$ has
to be added to the definition of the form factors. 

\subsection{Spectral statistics for the Gaussian random matrix ensembles}

We will now summarize the
relevant results from random-matrix
theory (for more details
see \cite{Guhr, chiral, Altland, Zirnbauer, Mehta, Porter}).

\subsubsection{The Wigner-Dyson ensembles}

The ergodic universality classes for
quantum systems in the Wigner-Dyson classes
are described by the well-known
Gaussian ensembles of random matrix theory GUE ($A$-GE), GOE
($A$I-GE) and GSE ($A$II-GE).  The universal spectral statistics is
invariant under shifts of the energy $\epsilon\rightarrow
\epsilon+\epsilon_0$. Thus the expectation value of the fluctuating
part of the density of states vanishes and so does its Fourier
transform
\begin{equation}
  K_1(\tau)^{\mathrm{W.D.}}=0.
\end{equation}  
The two-point correlation functions are given by
\begin{equation}
  \begin{array}{ll}
    \scriptstyle
    C^{\scriptscriptstyle \text{GUE}}(\epsilon)=&
    \scriptstyle\delta(\epsilon)- \frac{\sin^2 \pi
      \epsilon }{\pi^2 \epsilon^2}
    \\
    \scriptstyle C^{\scriptscriptstyle \text{GOE}}(\epsilon) =& 
    \scriptstyle C^{\scriptscriptstyle
      \text{GUE}}(\epsilon)+
    \frac{\left(\pi |\epsilon| \cos \pi \epsilon - \sin \pi |\epsilon|\right)
      \left(2\, \mathrm{Si}(\pi |\epsilon|)-\pi\right)
    }{2 \pi^2 \epsilon^2}
    \\
    \scriptstyle C^{\scriptscriptstyle \text{GSE}}(\epsilon) =&\scriptstyle
    \frac{1}{2}C^{\scriptscriptstyle \text{GUE}}(2\epsilon)
    +\frac{2 \pi |\epsilon| \cos 2 \pi \epsilon - \sin 2 \pi
      |\epsilon|}{4 \pi^2 \epsilon^2} \mathrm{Si}(2\pi|\epsilon|)
  \end{array}
\end{equation}
where $\mathrm{Si}(x)=\int_0^x \intd \xi\,  \frac{\sin \xi}{\xi}$ is the
sine integral.  The corresponding second-order form factors are given by
\begin{equation}
  \begin{array}{ll}
    \scriptstyle K_2^{\scriptscriptstyle \text{GUE}}(\tau)=&\scriptstyle
    \begin{cases}
      \scriptstyle |\tau| & \scriptstyle  \text{for $|\tau| < 1$}\\
      \scriptstyle 1 & \scriptstyle  \text{for $|\tau| \ge 1$}
    \end{cases}
    \\
    \scriptstyle K_2^{\scriptscriptstyle \text{GOE}}(\tau)=&\scriptstyle
    \begin{cases}
      \scriptstyle |\tau|\left(2 -\log(2|\tau|+1)\right)& \scriptstyle 
      \text{for $|\tau| < 1$}\\
      \scriptstyle 2-|\tau| \log\frac{2|\tau|+1}{2|\tau|-1} &  \scriptstyle 
      \text{for $|\tau| \ge
        1$}
    \end{cases}
    \\
    \scriptstyle K_2^{\scriptscriptstyle \text{GSE}}(\tau)= &\scriptstyle
    \begin{cases}
       \scriptstyle \frac{|\tau|}{4}\left(2 -\log||\tau|-1|\right)& \scriptstyle 
       \text{for $|\tau| < 2$}\\
      \scriptstyle  1& \scriptstyle  \text{for $|\tau| \ge 2$}
    \end{cases}.
  \end{array}
\end{equation}

\subsubsection{The novel ensembles}

The Gaussian random-matrix ensembles in the chiral symmetry classes
are known as chGUE ($A$III-GE), chGOE ($BD$I-GE) and chGSE ($C$II-GE).
The Andreev ensembles $C$-GE, $C$I-GE, $D$-GE, and $D$III-GE do not
have any established name.  The spectral statistics of these ensembles
is not invariant under energy-shifts and, as a consequence, deviations
from Weyl's law need not vanish near $\epsilon=0$. At energies much
larger than the mean level spacing $|\epsilon_0| \gg 1$ Wigner-Dyson
statistics is recovered. Thus, for the two-point correlation function
we have
\begin{equation}
  \begin{array}{lll}
    \scriptstyle
    C^{\scriptscriptstyle \text{chGUE}}(\epsilon,\epsilon_0),
    C^{\scriptscriptstyle \text{$C$-GE}}(\epsilon,\epsilon_0),
    C^{\scriptscriptstyle \text{$D$-GE}}(\epsilon,\epsilon_0)
    &\scriptstyle
    \xrightarrow{\epsilon_0 \gg 1}& 
    \scriptstyle C^{\scriptscriptstyle \text{GUE}}(\epsilon)\\
    \scriptstyle
    C^{\scriptscriptstyle \text{chGOE}}(\epsilon,\epsilon_0),
    C^{\scriptscriptstyle \text{$C$I-GE}}(\epsilon,\epsilon_0)
    & \scriptstyle
    \xrightarrow{\epsilon_0 \gg 1}& \scriptstyle 
    C^{\scriptscriptstyle \text{GOE}}(\epsilon)\\
    \scriptstyle C^{\scriptscriptstyle \text{chGSE}}(\epsilon,\epsilon_0),
    \scriptstyle C^{\scriptscriptstyle 
      \text{$C$I-GE}}(\epsilon,\epsilon_0) &
    \scriptstyle
    \xrightarrow{\epsilon_0 \gg 1}
    & \scriptstyle C^{\scriptscriptstyle \text{GSE}}(\epsilon).
  \end{array}
\end{equation}
The universal features near the symmetry point $\epsilon=0$ are most
prominent in the density of states.  Though there are universal
deviations from Wigner-Dyson statistics in all correlation functions
we will focus on the density of states.  The universal deviations from
Weyl's law for the chiral ensembles are given by
\cite{Mehta,chiral,Altland,dens1,chiraldensity,ivanov}
\begin{equation}
  \begin{array}{lll}
    \scriptstyle 
    \langle \delta d^{\scriptscriptstyle \text{chGUE}}(\epsilon) \rangle&
    \scriptstyle =&
    \scriptstyle 
    \frac{\pi^2 |\epsilon|}{2}\left({\scriptscriptstyle J_0^2(\pi\epsilon) +
      J_1^2(\pi\epsilon)}\right) -{\scriptscriptstyle 1}\\
    \scriptstyle 
    \langle \delta d^{\scriptscriptstyle \text{chGOE}}(\epsilon) \rangle&
    \scriptstyle =&
    \scriptstyle 
    \langle \delta d^{\scriptscriptstyle \text{chGUE}}(\epsilon)\rangle     
    +\frac{\pi}{2}{\scriptscriptstyle J_0(\pi \epsilon)}\left( 
      {\scriptscriptstyle 1-\int_0^{\pi|\epsilon|}
        d\xi J_0(\xi)}
    \right)\\
    \scriptstyle 
    \langle \delta d^{\scriptscriptstyle \text{chGSE}}(\epsilon) \rangle&
    \scriptstyle =& 
    \scriptstyle
    \langle \delta d^{\scriptscriptstyle \text{chGUE}}(2 \epsilon) \rangle
    -\frac{\pi}{2}{\scriptscriptstyle J_0(2\pi \epsilon)\,\int_0^{2\pi|\epsilon|} d\xi
    \, J_0(\xi)}
    \\
    \scriptstyle 
    \langle \delta d^{\scriptscriptstyle \text{$C$-GE}}(\epsilon) \rangle&
    \scriptstyle =&
    \scriptstyle 
    -\frac{\sin 2 \pi \epsilon}{2 \pi \epsilon}\\
    \scriptstyle 
    \langle \delta d^{\scriptscriptstyle \text{$C$I-GE}}(\epsilon) \rangle&
    \scriptstyle =&
    \scriptstyle 
    \langle \delta d^{\scriptscriptstyle \text{chGUE}}(\epsilon) \rangle
    -
    {\scriptscriptstyle \frac{\pi}{2}J_0(\pi\epsilon)J_1(\pi|\epsilon|)}\\
    \scriptstyle 
    \langle \delta d^{\scriptscriptstyle \text{$D$-GE}}(\epsilon) \rangle&
    \scriptstyle =&
    \scriptstyle -\langle \delta d^{\scriptscriptstyle \text{$C$-GE}}
    (\epsilon) \rangle
    \\
    \scriptstyle 
    \langle \delta d^{\scriptscriptstyle \text{$D$III-GE}}(\epsilon) \rangle&
    \scriptstyle =&
    \scriptstyle 
    \langle \delta d^{\scriptscriptstyle \text{$C$I-GE}}(2 \epsilon) \rangle+
    {\scriptscriptstyle \frac{\pi}{2}J_1(2\pi|\epsilon|)}
  \end{array}
\end{equation}

The corresponding first-order form factors can be calculated
explicitly in terms of the complete elliptic integrals of first,
second and third kind $\mathcal{K}(x)=\int_0^{\frac{\pi}{2}}
\frac{1}{\sqrt{1-x \sin^2 \phi}} d\phi $,
$\mathcal{E}(x)=\int_0^{\frac{\pi}{2}} \sqrt{1-x \sin^2 \phi}$ and
$\Pi(y,x)=\int_0^{\frac{\pi}{2}} \frac{1}{(1-y \sin^2\phi) \sqrt{1-x
    \sin^2 \phi}} d\phi$ (we use the convention that $\Pi(y,x)$ is
real for $y>1$ \cite{Stegun}). They are given by
\begin{equation}
  \begin{array}{lll}
    \scriptstyle K_1^{\scriptscriptstyle \text{chGUE}}(\tau)
    &\scriptstyle =
    &\scriptstyle \frac{|\tau|+1}{\pi
      |\tau|}\,\mathcal{E}
    \left(\frac{ 4|\tau|}{(1+|\tau|)^2}\right)-\\
    && \scriptstyle -
    \frac{1+\tau^2}{\pi |\tau|(1+|\tau|)}\,\mathcal{K} \left(\frac{
        4|\tau|}{ (1+|\tau|)^2}
    \right)\\
    \\
    \scriptstyle K^{\scriptscriptstyle \text{chGOE}}_1(\tau)
    &\scriptstyle =
    &\scriptstyle 
    K_1^{\scriptscriptstyle \text{chGUE}}(\tau)+ 
    \frac{ 1}{ \sqrt{1-4\tau^2}}
    \theta(1-2|\tau|)-\\
    &&\scriptstyle \frac{ 2|\tau|}{ \pi(|\tau|+1)(2|\tau|+1)}
    \Pi\left(\frac{4|\tau|}{ 2|\tau|+1}, \frac{ 4|\tau|}{
        (1+|\tau|)^2}\right)\\
    \\
    \scriptstyle K^{\scriptscriptstyle \text{chGSE}}_1(\tau)
    &\scriptstyle =
    &\scriptstyle 
    \frac{1}{2} K_1^{\scriptscriptstyle \text{chGUE}}(\frac{\tau}{2})
    -\frac{ 1}{2 \sqrt{1-\tau^2}} \theta(1-|\tau|)
    -
    \\
    &
    &\scriptstyle 
    \frac{ |\tau|}{\pi(|\tau|+2)(|\tau|+1)}
    \Pi\left(\frac{2|\tau|}{|\tau|+1}, \frac{ 8|\tau|}{
        (2+|\tau|)^2}\right)\\
    \\
    \scriptstyle K_1^{\scriptscriptstyle \text{$C$-GE}}(\tau)&\scriptstyle =&
    \scriptstyle -\theta(1-|\tau|)\\
    \\
    \scriptstyle K^{\scriptscriptstyle \text{$C$I-GE}}_1(\tau)&\scriptstyle =&
    \scriptstyle  \frac{|\tau|+1}{\pi|\tau|}
    \mathcal{E}\left(\frac{4|\tau|}{(|\tau|+1)^2}\right)
    +\frac{|\tau|-1}{\pi|\tau|}
    \mathcal{K}\left(\frac{4|\tau|}{(|\tau|+1)^2}\right)
    -1\\
    \\
    \scriptstyle 
    K^{\scriptscriptstyle \text{$D$-GE}}_1(\tau)&\scriptstyle =&
    \scriptstyle \theta(1-|\tau|)\\
    \\
    \scriptstyle K^{\scriptscriptstyle \text{$D$III-GE}}_1(\tau)&
    \scriptstyle =&
    \scriptstyle  
    \frac{1}{2}K_1(\frac{\tau}{2})^{\scriptscriptstyle \text{$C$I-GE}}
    +1-\theta(|\tau|-1)\frac{|\tau|}{\sqrt{\tau^2-1}}.
  \end{array}
\end{equation}

\subsection{\label{sec:BGS-conjecture} The generalization
  of the Bohigas-Giannoni-Schmit conjecture}

It has been conjectured by Bohigas, Giannoni and Schmit that
quantum systems (in the semiclassical regime) 
with a chaotic classical limit have
universal spectral fluctuations that coincide with the
predictions of one of the Wigner-Dyson Gaussian ensembles
of random-matrix theory GUE, GOE, or GSE. More precisely 
in an average over different parts of the unfolded
spectrum the
$n$-point correlation functions for $n\ge 2$ of a \emph{single}
spectrum are conjectured to coincide with the corresponding correlation
functions of the Wigner-Dyson ensemble. 
The mean density of states of a given quantum system
is non-universal and cannot be described by random-matrix theory.
Semiclassically it is given by Weyl's law.

A lot of evidence has since been gathered both numerical and analytical 
that this conjecture is true in generic chaotic systems 
\cite{Guhr} (though a few exceptions are 
known \cite{Keating,Bogomolny}). 
Many approaches have been used to understand and proof the fidelity
to random-matrix theory in complex quantum systems \cite{disordered, do1,do2}.
Recently there has been considerable progress
in the semiclassical approach using periodic orbit 
theory \cite{Sieber}.

Bohigas \textit{et al} stated their conjecture 
before the impact
of spectral mirror symmetries on spectral statistics has been 
recognized.
A proper generalization of their statement has to take into account that 
a spectral average will
wipe out all effects of a spectral mirror symmetry.
Thus the
original conjecture is expected to hold for the 
novel symmetry classes as well:
after averaging over different parts of a single spectrum
they will show the universal spectral
fluctuations of GUE, GOE, or GSE.

The additional universal features in physical 
systems near the spectral
symmetry point can only be observed when an average
over various spectra is performed. 
This corresponds to an average over some system parameter.
We conjecture that for classically chaotic
systems with a spectral symmetry 
\emph{all} correlation functions of the
fluctuating part of the unfolded density of states $\delta d(\epsilon)$
as given by \eqref{eq:unfolded_dos} averaged over \emph{one} system
parameter coincide with those of the corresponding
Gaussian random-matrix ensemble in the novel symmetry classes.
This includes universal
deviations from Weyl's law in the density of states itself.
Note, that though there are seven symmetry classes 
there are infinitely many ergodic universal classes due to the 
different subclasses. 
Though some average
is certainly necessary we may still conjecture the fidelity
to ergodic random-matrix theories of a single physical system
by formally averaging over spectra for different values of 
an effective Planck's constant. 
In superconducting-normalconducting hybrid 
structures this corresponds to an average over Fermi energy $\mu$.

\section{Quantum star graphs for the ten symmetry classes
  \label{sec:stargraphs}}

Quantum graphs have been introduced by Kottos and Smilansky 
\cite{Kottos} as simple
quantum systems with an exact semiclassical trace formula for the
density of states. They consist of $V$ vertices and connected
by $B$ bonds. 
Each bond $b_i$ connects two vertices and has a length $L_i$. A
particle propagates freely on the bonds and is scattered at the
vertices by prescribed boundary conditions which leads to
quantization. In their first approach
Kottos and Smilansky considered
vertex boundary conditions that implied 
current conservation and continuity of the wave function.
The continuity condition is not always essential and has often been 
relaxed. 
In that case the boundary conditions at a vertex are
specified by any unitary 
scattering matrix that transforms incoming waves to outgoing waves --
unitarity of the vertex scattering matrix
is equivalent to current conservation.

We will not discuss general graphs but limit ourselves
to a very simple class of graphs -- \emph{star graphs}.

\subsection{Quantization of star graphs}

\begin{figure}
  \begin{center}
    \includegraphics[width=0.35\linewidth]{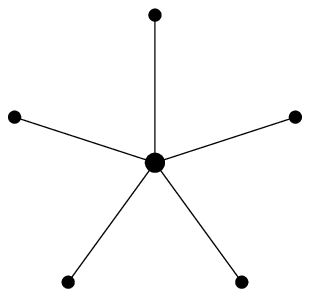}
    \caption{\label{fig:stargraph} Sketch of a star graph with
      five bonds of equal length.}
  \end{center}
\end{figure}

A star graph consists of $B$ bonds $b_j$ of length $L_j$
($j=1,\dots,B$) and $V=B+1$ vertices $v_j$ ($j=0,\dots,B$). Each bond
$b_j$ emanates from the central vertex $v_0$ and connects it to the
peripheral vertex $v_j$ (see figure \ref{fig:stargraph}).

We will allow for a multi-component wave function on
the graph. The number of components $\mu$ is assumed to be equal on
all bonds. It may represent different spin
components or electron and hole
components of a quasi-particle.  
The $\mu$-component wave function on the bond $b_j$ is
\begin{equation}
  \boldsymbol{\Psi}^{(j)}(x^{(j)})= 
  \boldsymbol{\phi}_{\mathrm{out}}^{(j)} \,
  \Exp^{\ImgI k x^{(j)}}+
  \boldsymbol{\phi}_{\mathrm{in}}^{(j)}\, 
  \Exp^{\ImgI k (L_j-x^{(j)})}
\end{equation}
where $0 \le x^{(j)} \le L_j$ is the distance from the central vertex
and
\begin{equation}
  \boldsymbol{\phi}_{\mathrm{in\, (out)}}^{(j)}=
  \begin{pmatrix}
    \scriptstyle\phi_{\mathrm{in\,(out)},1}^{(j)}\\
    \scriptstyle\phi_{\mathrm{in\, (out)},2}^{(j)}\\
    \scriptstyle\dots\\
    \scriptstyle\phi_{\mathrm{in\,(out)},\mu}^{(j)}
  \end{pmatrix}
\end{equation}
are $\mu$-component vectors of constant coefficients 
for the incoming (outgoing) waves on bond $b_j$
(``incoming'' and ``outgoing'' will always be used with respect 
to the central vertex).

It is convenient to combine all the coefficients of incoming 
(outgoing) waves into two vectors of dimension $\mu B$
\begin{equation}
  \boldsymbol{\phi}_{\mathrm{in\, (out)}}=
  \begin{pmatrix}
    \scriptstyle \phi_{\mathrm{in\, (out)},1}^{(1)}\\
    \scriptstyle\dots\\
    \scriptstyle\phi_{\mathrm{in\, (out)},1}^{(B)}\\
    \scriptstyle\dots\\
    \scriptstyle\phi_{\mathrm{in\, (out)},\mu}^{(1)}\\
    \scriptstyle\dots\\
    \scriptstyle\phi_{\mathrm{in\, (out)},\mu}^{(B)}
  \end{pmatrix}.
\end{equation}
The boundary condition at the center can then be written in the form
\begin{equation}
  \boldsymbol{\phi}_{\mathrm{out}}=\mathcal{S}_C \mathcal{L}(k) \, 
  \boldsymbol{\phi}_{\mathrm{in}}.
  \label{eq:bc_center}
\end{equation}  
The diagonal $\mu B \times \mu B$ matrix
\begin{equation}
  \mathcal{L}_{\alpha j, \alpha' j'}(k)=\delta_{j j'} 
  \delta_{\alpha, \alpha'}
  \Exp^{\ImgI k L_j}
\end{equation}
describes the propagation along the bonds.
Here (and in the rest of this section) 
$\alpha=1,2,\dots \mu$ indicates the component of the wave function
and $j=1,2,\dots,B$ is the bond index. 
The 
\emph{central scattering matrix} $\mathcal{S}_C$ is 
a fixed  unitary $\mu B \times \mu B$ matrix that
defines the boundary conditions at the center. 
For definiteness, we will assume that 
different components of the wave
function do not mix at the center, thus
\begin{equation}
  \mathcal{S}_{C,\alpha j , \alpha' j'}=\delta_{\alpha \alpha'} 
  \mathcal{S}_{C, j j'}^{(\alpha)} = \delta_{\alpha \alpha'}
  a^{(\alpha)}_{C, j j'} \Exp^{\ImgI w^{(\alpha)}_{C, j j'}}
\end{equation} 
where the $B \times B$ matrix $S_{C}^{(\alpha)}$ describes scattering
of the $\alpha$-component at the center.  

The boundary conditions at
the peripheral vertices may be described by one
fixed $\mu \times \mu$
vertex scattering matrices for each peripheral vertex  -- 
these can be combined
to a single unitary $\mu B \times \mu B$ peripheral scattering matrix
$\mathcal{S}_P$ such that
\begin{equation}
  \boldsymbol{\phi}_{\mathrm{in}}=\mathcal{S}_P \mathcal{L}(k)  \,
  \boldsymbol{\phi}_{\mathrm{out}}.
  \label{eq:bc_rand}
\end{equation}

Since different bonds are not coupled at the peripheral vertices
\begin{equation}
  \mathcal{S}_{P,\alpha j, \alpha' j'}=\delta_{j j'} 
  \sigma^{(j)}_{\alpha \alpha'}=
  \delta_{j j'} a^{(j)}_{P, \alpha \alpha'} 
  \Exp^{\ImgI w^{(j)}_{P, \alpha \alpha'}} 
\end{equation}
where $\sigma^{(j)}$ is the $\mu \times \mu$ scattering matrix at the
vertex $v_j$.  

We will not allow any dependence of
the scattering matrices  $\mathcal{S}_C$ and $\mathcal{S}_P$
on the wave number $k$.
Uniqueness of the wave function and the boundary
conditions \eqref{eq:bc_center} and \eqref{eq:bc_rand} lead to
the quantization condition
\begin{equation}
  \boldsymbol{\phi}_{\mathrm{in}}=
  \mathcal{S}_P \mathcal{L}(k)\mathcal{S}_C \mathcal{L}(k)  \,
  \boldsymbol{\phi}_{\mathrm{in}}
  \equiv \mathcal{S}_B(k)\, \boldsymbol{\phi}_{\mathrm{in}}
  \label{eq:qc1}
\end{equation}
where we introduced the \emph{bond scattering matrix}
$\mathcal{S}_B(k)$.
Non-trivial solutions of these equations exist only when
the wave number belongs to the discrete spectrum $k=k_n$
given by the zeros of the corresponding determinant
\begin{equation}
  \mathrm{det}\left( \mathcal{S}_B(k_n)- \ONE \right)=0.
  \label{eq:quantization_condition}
\end{equation}
The density of states for the graph is defined as
\begin{equation}
  d(k)=\frac{1}{g}\sum_{n=0}^{\infty} \delta (k-k_n),
  \label{eq:graph-DOS}
\end{equation}
where $g=2$ in systems with Kramers' degeneracy (else $g=1$).

\subsection{The trace formula}

Let us now write the density of states as a sum of its mean
$d_{\mathrm{Weyl}}$ and an oscillating part $\delta d(k)$
\begin{equation}
  d(k)=d_{\mathrm{Weyl}}+ \delta d(k).
\end{equation} 
For both contributions one can give an exact semiclassical expression.
The mean density of states is given by Weyl's law
\begin{equation}
  d_{\mathrm{Weyl}}=\frac{\mu \sum_j L_j}{g \pi}
\end{equation}
and the oscillating part obeys the trace formula \cite{Kottos}
\begin{equation}
  \delta d(k)= \IM \frac{d}{dk} \sum_{n=1}^\infty \frac{1}{g \pi\,n}
  \,\mathrm{tr}\, \mathcal{S}_B(k)^n.
  \label{eq:trace1}
\end{equation}
In the sequel we will consider star
graphs where all bond lengths are equal $L_j=L$. 
In that case the bond
scattering matrix is a periodic function of $k$
\begin{equation}
  \mathcal{S}_B(k)=\mathcal{S}_B(k+\frac{\pi}{L})=\Exp^{\ImgI 2 k L}
  \tilde{\mathcal{S}}_B
\end{equation}
where
\begin{equation}
  \tilde{\mathcal{S}}_B=\mathcal{S}_B(k=0)=\mathcal{S}_P \mathcal{S}_C
\end{equation}
is the \emph{reduced bond scattering matrix}.  Thus, the spectrum is
also periodic and the trace formula simplifies to
\begin{equation}
  d(k)=\frac{\mu B \,L}{g \pi}+ \frac{2 L}{g \pi}\, \RE \sum_{n=1}^\infty
  \Exp^{\ImgI 2 n k L}\, \mathrm{tr}\, {\tilde{\mathcal{S}}_B}^n.
  \label{eq:trace2}
\end{equation}
The periodicity of the spectrum will not be relevant here
as we are interested in features on the scale
of a mean level spacing.

For equal bond lengths the trace formula can be derived in a few
lines: Let $\Exp^{-\ImgI \phi_j}$ ($j=1,\dots,\mu B$) be the
eigenvalues of the unitary reduced bond scattering matrix
$\tilde{\mathcal{S}}_B$. The quantization condition
\eqref{eq:quantization_condition} is equivalent to $k=\frac{\phi_j}{2
  L}\, \mathrm{mod}\,\frac{\pi}{L}$ ($j=1,\dots, \mu B$) and the
density of states is
\begin{equation}
  \begin{split}
    d(k)&=\frac{1}{g}\sum_{j=1}^{\mu B} \sum_{n=-\infty}^{\infty}
    \delta(k-\frac{\phi_j}{2L}+n \frac{\pi}{L})\\
    &=\frac{L}{g \pi}\sum_{n=-\infty}^{\infty} \Exp^{\ImgI 2 n k L }
    \sum_{j=1}^{\mu B} \Exp^{-\ImgI n \phi_j }.
  \end{split}
\end{equation}
The mean density of states $d_{\mathrm{Weyl}}=\frac{\mu B L}{g \pi}$ is
just the $n=0$ term in the sum over $n$ while the rest gives the trace
formula for the oscillating part (the second line follows from the
first by Poisson's summation formula $\sum_{n=-\infty}^\infty
\delta(x-n)=\sum_{n=-\infty}^\infty \Exp^{\ImgI 2 \pi n x}$).

The trace formula \eqref{eq:trace1} can be interpreted as a sum over
\emph{periodic orbits} $p$ on the graph. A periodic orbit
$p=[ (j_1,\alpha_{1}),(j_2,\alpha_{2}),\dots (j_n,
    \alpha_{n})]$
of length $n$ is defined by a sequence of $n$ peripheral vertices
$v_{j_1}v_{j_2}\dots v_{j_n}$ visited one after the
other together with the specification
of the wave component $\alpha_j$ between two vertices (cyclic
permutations define the same orbit). 
A periodic orbit is \emph{primitive} if it is not
the repetition of a shorter periodic orbit. In terms of primitive
periodic orbits $p$ and its repetitions the trace formula reads
\begin{equation}
  \delta d(k) = \frac{2 L}{g \pi}\sum_{\textrm{p.p.o.:} p}\sum_{r=1}^\infty
  n_p \left({A_p} \Exp^{\ImgI  W_p}\right)^r
  \label{eq:trace_po}
\end{equation} 
where $n_p$ is the length of the primitive periodic orbit
$A_p=\prod_{l=1}^{n_p} {a_P}^{(j_{l+1})}_{\alpha_{{l+1}}\alpha_{{l}}}
{a_C}^{(\alpha_{l})}_{j_{l+1} j_{l}}$
is the amplitude of  the primitive orbit and
$W_p= 2 n_p L k + \sum_{l=1}^{n_p} \left(
    {w_C}^{(\alpha_l)}_{ j_{l+1} j_{l}}+
    {w_P}^{(j_{l+1})}_{\alpha_{{l+1}}\alpha_{{l}}}
  \right)$
its phase (``action''). Note, that we set $j_{n_p+1}=j_1$ and
$\alpha_{n_p+1}=\alpha_1$.
  
The similarity of the sum over periodic
orbits \eqref{eq:trace_po} to the semiclassical Gutzwiller trace
formula is evident. However, while semiclassics, in general, is an
approximation the semiclassical trace formula for quantum graphs is
exact.

The trace formula will be our main tool in the analysis of universal
spectral statistics. It will lead
us to a simple expression for the form factors that
can easily be averaged numerically.
In the second paper of this series the trace formula will be in the
center of an analytic approach to universality.

Since universality
exists on the scale of the mean level spacing we will write $k=
\kappa \Delta k$ where $\Delta k=\frac{g \pi}{\mu B L}$ is the mean
level spacing. In terms of the rescaled wave number the trace formula
is
\begin{equation}
  d(\kappa) = 1+ \frac{2}{\mu B}\, \RE \sum_{n=1}^{\infty}
  \Exp^{\ImgI 2 \pi \kappa \frac{ g n}{\mu B}} s_n.
  \label{eq:trace3}
\end{equation}
We have introduced the shorthand
\begin{equation}
  s_n=\mathrm{tr}\, 
  {\tilde{\mathcal{S}}_B}^n
  \label{eq:trace_sn}
\end{equation}
for the $n$-th trace of the reduced bond scattering matrix.

The first-order form factor is obtained
by a Fourier transform and a subsequent time average.
It obeys the trace formula
\begin{equation}
  K_1(\tau)=
  \frac{g}{\mu B}
  \overline{
    \sum_{n=1}^\infty\delta\left(|\tau| - \frac{g n}{\mu B}\right) K_{1,n}}
\end{equation}
where the bar denotes a time average over a small time interval 
$\Delta \tau =\frac{g \Delta n}{\mu B}\ll 1$ and
\begin{equation}
  K_{1,n}= \frac{2}{g}\, \left\langle s_n
  \right\rangle.
\end{equation}
The brackets $\langle \cdot \rangle$ denote an average
over an ensemble of graphs.  
This can be written more compactly as
\begin{equation}
  K_1(\tau\equiv \frac{g n}{\mu B})=\overline{K_{1,n}}\equiv
  \frac{1}{\Delta n}\sum_{k=0}^{\Delta n -1} K_{1,n+k}
\end{equation}
where the continuous time average has been replaced by
an average over the discrete time $\tau\equiv\frac{g n}{\mu B}$.

The second-order form factor for a graph also obeys a trace formula
which, after a spectral average over the central wave number, is given
by
\begin{equation}
  K_2(\tau\equiv \frac{g n}{\mu B})
  =\overline{K_{2,n}}
  \label{eq:2nd_formfactor_trace}
\end{equation}
where
\begin{equation}
  K_{2,n}=\frac{1}{g \mu B} 
  \left\langle \left|s_n\right|^2 \right\rangle.
\end{equation}
If no spectral average is performed additional terms
appear. These are irrelevant for the
graphs in the Wigner-Dyson classes (they do not survive the
subsequent
ensemble average). Here, we will not consider the second-order 
form factor
for graphs in the novel symmetry classes where the additional terms 
are relevant near the central energy $\epsilon_0=0$.

Though $K_{1,n}$ and $K_{2,n}$ do not involve a time average we will 
refer to them as (discrete time) form factors.

\subsection{Star graphs for all symmetry classes}

We will now construct ensembles of star graphs for each symmetry
class. The star graphs will be constructed in such a way that
spectral fluctuations of the corresponding ergodic universality
classes can be expected. 
Though we are not able to prove 
an equivalent conjecture we will give strong evidence.

The constructions of star graphs for
each symmetry class are based on a proper choice of the central
and peripheral scattering matrices $\mathcal{S}_C$ and
$\mathcal{S}_P$. Both have to obey the right symmetry conditions
(see section \ref{sec:symmetry}). 

As an example of star graphs in class $A$I
that \emph{do not} belong to the  ergodic universality class let
us mention \emph{Neumann} star graphs.
These have a one-component wave
function ($\mu=1$) on the bonds, Dirichlet boundary conditions at the
peripheral vertices such that $\mathcal{S}_P=-\ONE$ and Neumann
boundary conditions at the center, thus
$\mathcal{S}_{C,kl}=\frac{2}{B}-\delta_{kl}$.  Such graphs have been
investigated first in \cite{Kottos} and in more 
detail in \cite{gregonstars} -- in contrast to our approach 
the bond lengths were chosen different for each bond (and
incommensurate). However, Neumann boundary conditions at the center
favor backscattering and lead to non-universal (localization)
effects \cite{gregonstars}.

Our approach is different in as
much as we allow for more general scattering matrices at the center
and in as much as we will always consider an ensemble of graphs.  
The occurrence or non-occurrence
of localization effects can be traced back to
a gap condition on the matrix 
$\mathcal{T}_{ij}\equiv|\mathcal{S}_{C;ij}|^2$.
This bistochastic matrix describes the corresponding ``classical''
probabilistic dynamics (equivalent to a Frobenius-Perron
operator). In chaotic (ergodic) systems 
the Frobenius-Perron
operator has a finite gap in the spectrum between the (unique)
eigenvalue one and all other eigenvalues which describe the decay
of the probability distribution.
In Neumann star graphs this gap is small and vanishes
in the limit $B\rightarrow \infty$ faster than $\frac{1}{B}$
which leads to non-ergodic spectral statistics.

In general one needs a multi-component wave
function to introduce the different symmetries.  
The number of components $\mu$ has
been chosen minimal under the additional assumptions that the
components do not mix at the central vertex and that time-reversal is
only broken at the peripheral vertices. 

Though we explicitly choose the
central and peripheral scattering matrices
guided by simplicity and minimality, most of
the results are much more general. 

The central scattering can be chosen in a very simple way by using the
symmetric $B \times B$ \emph{discrete Fourier transform matrix}
\cite{Tanner}
\begin{equation}
  \mathcal{S}_{\mathrm{DFT},kl}=
  \frac{1}{\sqrt{B}}\Exp^{\ImgI 2\pi \frac{kl}{B}}
\end{equation}
or its complex conjugate for each component.
An incoming wave on a given bond is scattered with equal
probability to any bond which excludes localization effects.
Indeed, the matrix 
$\mathcal{T}_{\mathrm{DFT},kl}=|\mathcal{S}_{\mathrm{DFT},kl}|^2=\frac{1}{B}$
has one eigenvalue $1$ while all other $B-1$
eigenvalues vanish (so the gap is maximal). 

The bond
scattering matrix $\mathcal{S}_B(k)=\mathcal{S}_C
\mathcal{L}(k)\mathcal{S}_P\mathcal{L}(k)$ for each ensemble of graphs
is constructed by demanding that the matrices $\mathcal{S}_C$,
$\mathcal{S}_P$, and $\mathcal{L}(k)$ do all have the canonical forms
of the desired symmetry class given in section  \ref{sec:symmetry}.
Note, that for star graphs we are interested in the $k$ spectrum, so
in the canonical forms for scattering matrices the energy $E$ has 
to be replaced by $k$.  The
ensemble of graphs is built by introducing some random phases into the
peripheral scattering matrix.

\subsubsection{Star graphs in the Wigner-Dyson classes}

Let us start with the simplest case: an ensemble of star graphs in
class $A$I where a one-component wave-function suffices to incorporate
the time-reversal symmetry which demands 
that the unitary matrices $\mathcal{S}_C$, $\mathcal{S}_P$, and
$\mathcal{L}$ are all symmetric. 
Now, $\mathcal{S}_P$ and
$\mathcal{L}(k)=\Exp^{\ImgI k L}\ONE$ are diagonal for $\mu=1$, and
choosing
\begin{equation}
  \text{$A$I}:\qquad \mathcal{S}_C= \mathcal{S}_{\mathrm{DFT}}
\end{equation}
we meet all requirements. At the peripheral vertices we are free to
choose one random phase $\beta_k$ for each peripheral
vertex $j$ independently such that
\begin{equation}
  \text{$A$I}:\qquad \mathcal{S}_{P,kl}=  \delta_{kl}\Exp^{\ImgI \beta_k} 
\end{equation}
where $0\le \beta_k<2 \pi$ is uniformly distributed.

\begin{figure}
  \begin{center}
    \includegraphics[width=0.8\linewidth]{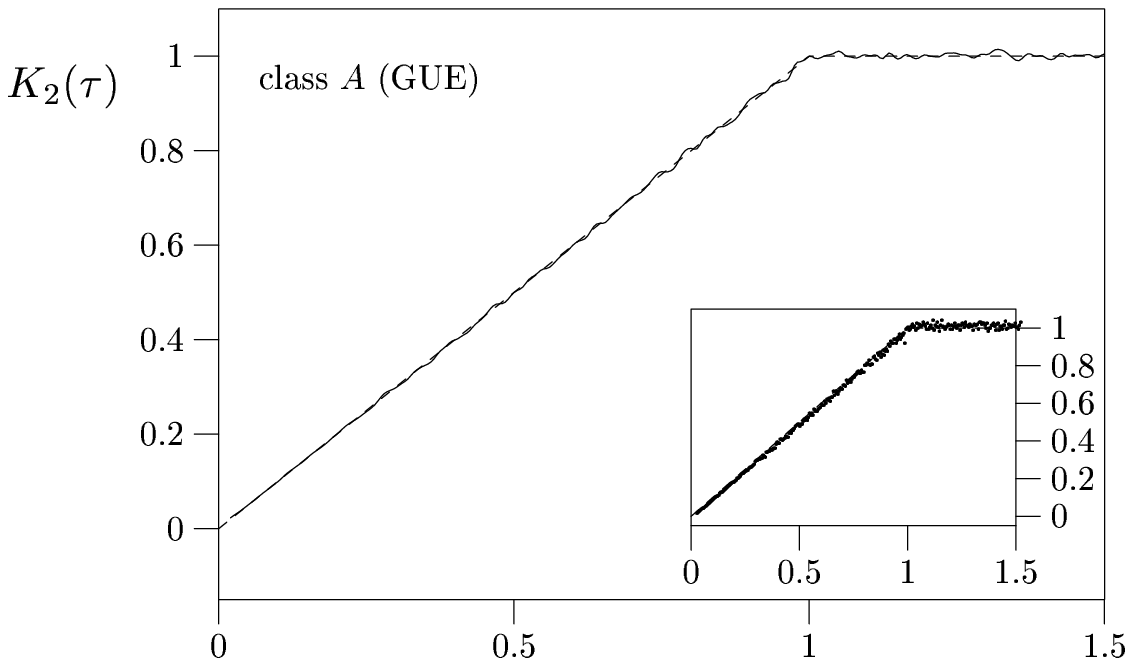}
    \includegraphics[width=0.8\linewidth]{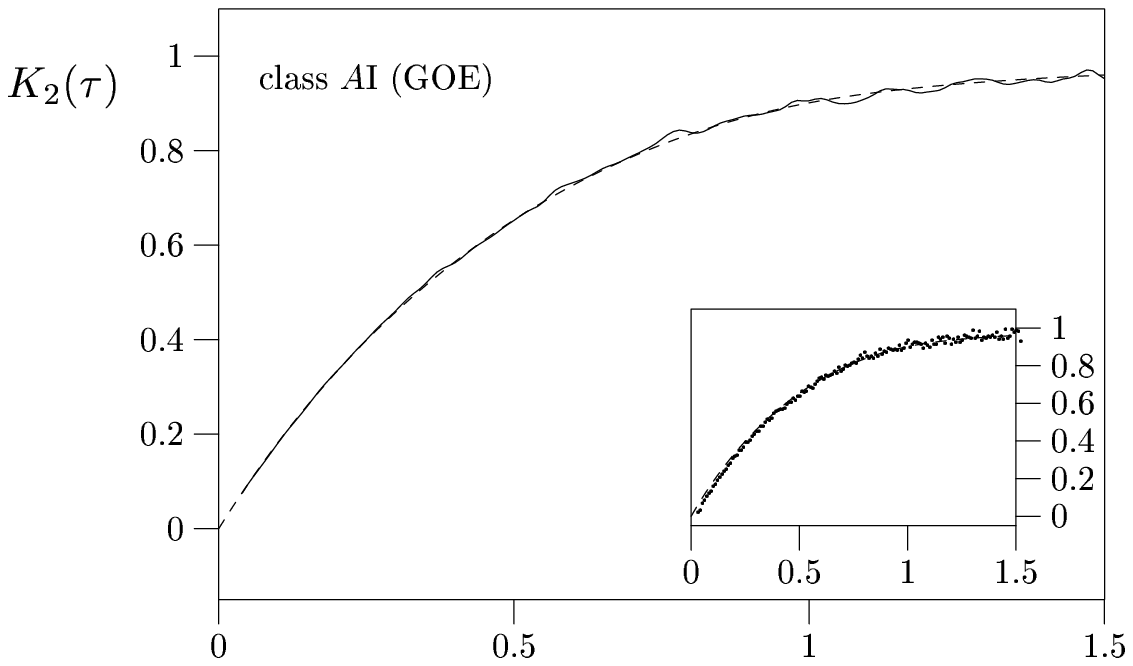}
    \includegraphics[width=0.8\linewidth]{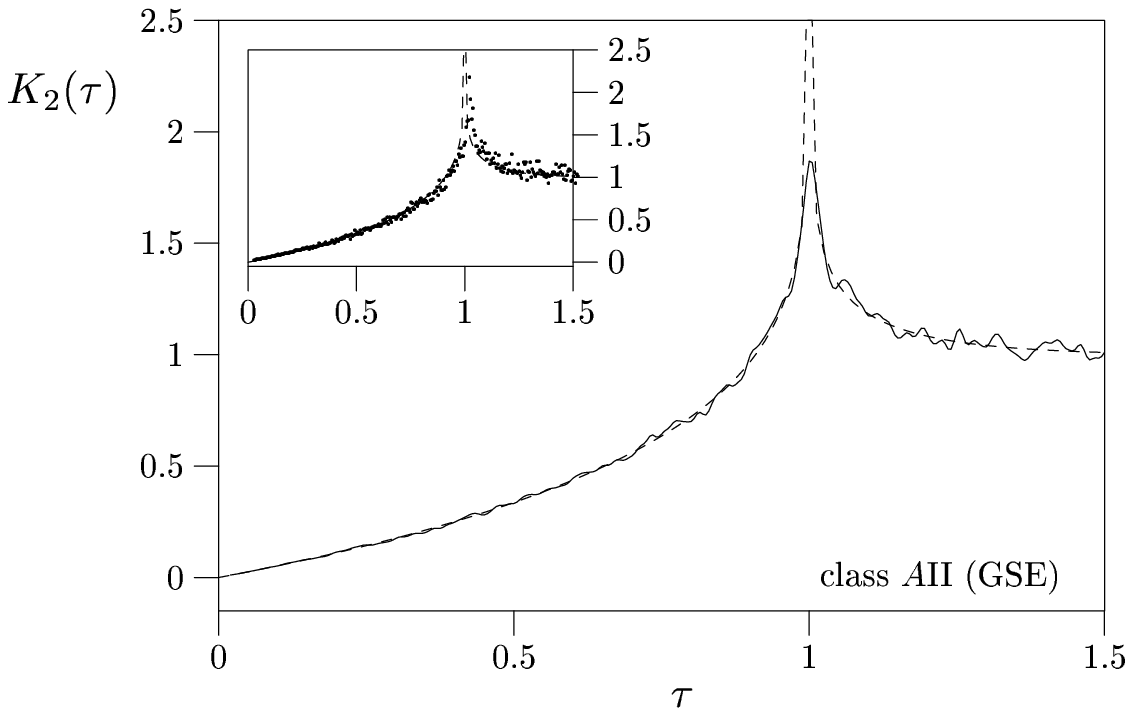}
    \caption{\label{fig:numWD} Second-order form factor
      for star graphs in the Wigner-Dyson ensembles
      averaged over 
      $10 000$ realizations of with $B=100$ bonds
      (additional time average over an interval
      of length $8 \frac{g}{\mu B}$). 
      Dashed lines: prediction by Gaussian random-matrix theory. 
      Full lines: numerically calculated form factor for graphs.
      Inlets: discrete time form factor $K_{2,n}$
      as function of
      $\tau=n \frac{g}{\mu B}$
      for the class $A$ and $A$I graphs. 
      For the class $A$II graphs $K_{2,n}$
      vanishes for odd $n$ -- the inlet shows $\frac{K_{2,n}}{2}$
      for even $n$.}
  \end{center}
\end{figure}

For class $A$ we have to break time-reversal symmetry. This may be
done by choosing a non-symmetric central scattering matrix for a
one-component wave-function on the graph. As we like to keep the
simplicity of the discrete Fourier transform matrix we choose another
simple construction with
a two-component wave function, a central 
$2B \times 2B$ scattering matrix
\begin{equation}
  \text{$A$}:\qquad
  \mathcal{S}_C=
  \begin{pmatrix}
    \mathcal{S}_{\mathrm{DFT}} &0 \\
    0 & \mathcal{S}_{\mathrm{DFT}}
  \end{pmatrix},
\end{equation}
and
\begin{equation}
  \text{$A$}:\qquad
  \mathcal{S}_P=
  \frac{1}{\sqrt{2}}
  \begin{pmatrix}
    \mathcal{D}_1 & \mathcal{D}_2\\
    \mathcal{D}_3 & \mathcal{D}_4
  \end{pmatrix}
\end{equation}
for the peripheral scattering matrix. The diagonal matrices
$\mathcal{D}_j$ are
\begin{equation}
  \text{$A$}:\qquad
  \begin{array}{rl} 
    \mathcal{D}_{1,kl} &= \delta_{kl} \Exp^{ \ImgI (\beta_k+\gamma_k)}\\
    \mathcal{D}_{2,kl} &= \delta_{kl} \ImgI \Exp^{\ImgI (\beta_k+\delta_k) }\\
    \mathcal{D}_{3,kl} &= \delta_{kl} \ImgI \Exp^{\ImgI(\beta_k-\delta_k)}\\
    \mathcal{D}_{4,kl} &= \delta_{kl} \Exp^{\ImgI(\beta_k-\gamma_k)}
  \end{array}
\end{equation}
where the independent random phases $\beta_k$, $\gamma_k$, and
$\delta_k$ are uniformly distributed.

For class $A$II a four-component wave-function is needed to
incorporate time-reversal invariance with $\mathcal{T}^2=-\ONE$ into
our scheme for star graphs. Indeed, the number of components must be
even as discussed above in \ref{sec:Wigner-Dyson}. In addition, we
assumed that components do only mix at the peripheral vertices. Then, a
$4\times 4 $ scattering matrix at the peripheral vertices is the
minimal matrix dimension that allows for component mixing as can be
seen from the canonical form \eqref{eq:canon_S_AII} of a $A$II
scattering matrix.  The diagonal matrix $\mathcal{L}(k)$ is a diagonal
unitary matrix of the canonical form.  All further requirements are
met by choosing
\begin{equation}
  \text{$A$II}:\qquad
  \mathcal{S}_C=
  \begin{pmatrix}
    \mathcal{S}_{\mathrm{DFT}} &0 &0&0\\
    0 & \mathcal{S}_{\mathrm{DFT}}&0&0\\
    0&0&\mathcal{S}_{\mathrm{DFT}}&0\\
    0&0&0&\mathcal{S}_{\mathrm{DFT}}
  \end{pmatrix}
\end{equation}
for the central $4B \times 4B$ scattering matrix, and
\begin{equation}
  \text{$A$II}:\qquad
  \mathcal{S}_P=
  \frac{1}{\sqrt{2}}
  \begin{pmatrix}
    0&\mathcal{D}_1&0&\mathcal{D}_2\\
    \mathcal{D}_3 &0& -\mathcal{D}_2&0\\
    0&-\mathcal{D}_4&0&\mathcal{D}_3\\
    \mathcal{D}_4 &0& \mathcal{D}_1&0
  \end{pmatrix}
\end{equation}
for the peripheral scattering matrix. The diagonal matrices
$\mathcal{D}_j$ are given by
\begin{equation}
  \text{$A$II}:\qquad
  \begin{array}{rl}
    \mathcal{D}_{1,kl} &= \delta_{kl} \Exp^{ \ImgI (\beta_k+\gamma_k)}\\
    \mathcal{D}_{2,kl} &= \delta_{kl} \Exp^{\ImgI (\beta_k+\delta_k) }\\
    \mathcal{D}_{3,kl} &= \delta_{kl} \Exp^{\ImgI(\beta_k-\gamma_k)}\\
    \mathcal{D}_{4,kl} &= \delta_{kl} \Exp^{\ImgI(\beta_k-\delta_k)},
  \end{array}
\end{equation}
where the independent random phases $\beta_k$, $\gamma_k$, and
$\delta_k$ are uniformly distributed. 

Ergodic spectral statistics may be expected
for the three ensembles in The Wigner-Dyson classes.
This is strongly supported by a numerical calculation of
the second-order form
factor (see figure \ref{fig:numWD}). 

\subsubsection{\label{sec:chiralstars} Chiral and Andreev star graphs}

\begin{figure}
  \begin{center}
    \includegraphics[width=0.8\linewidth]{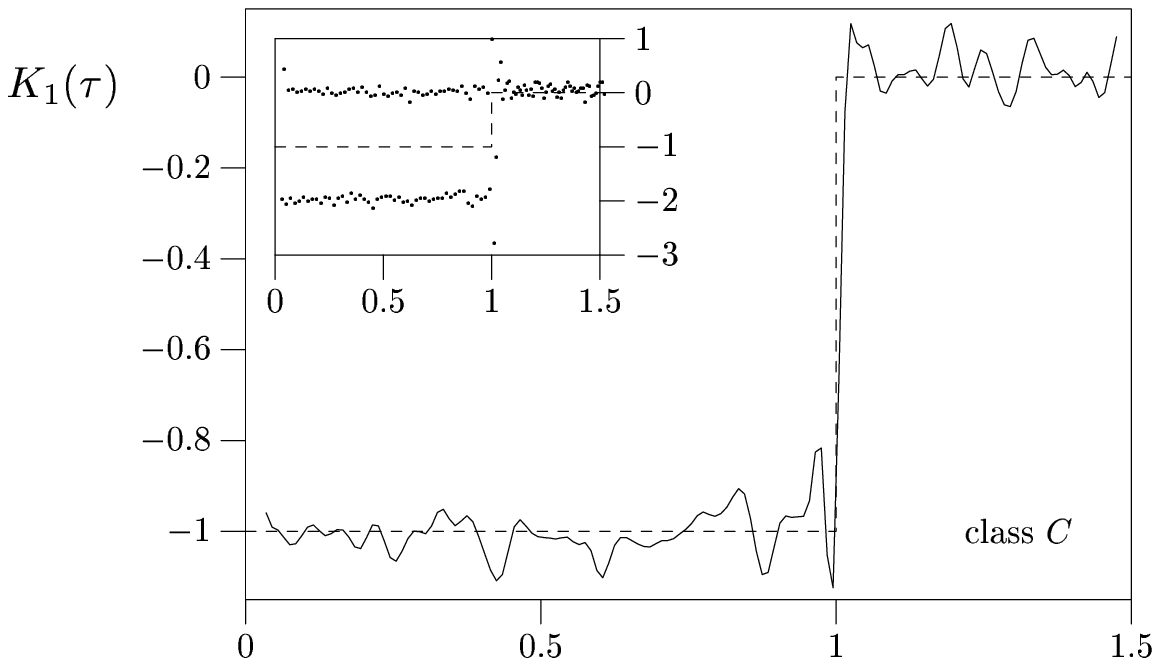}
    \includegraphics[width=0.8\linewidth]{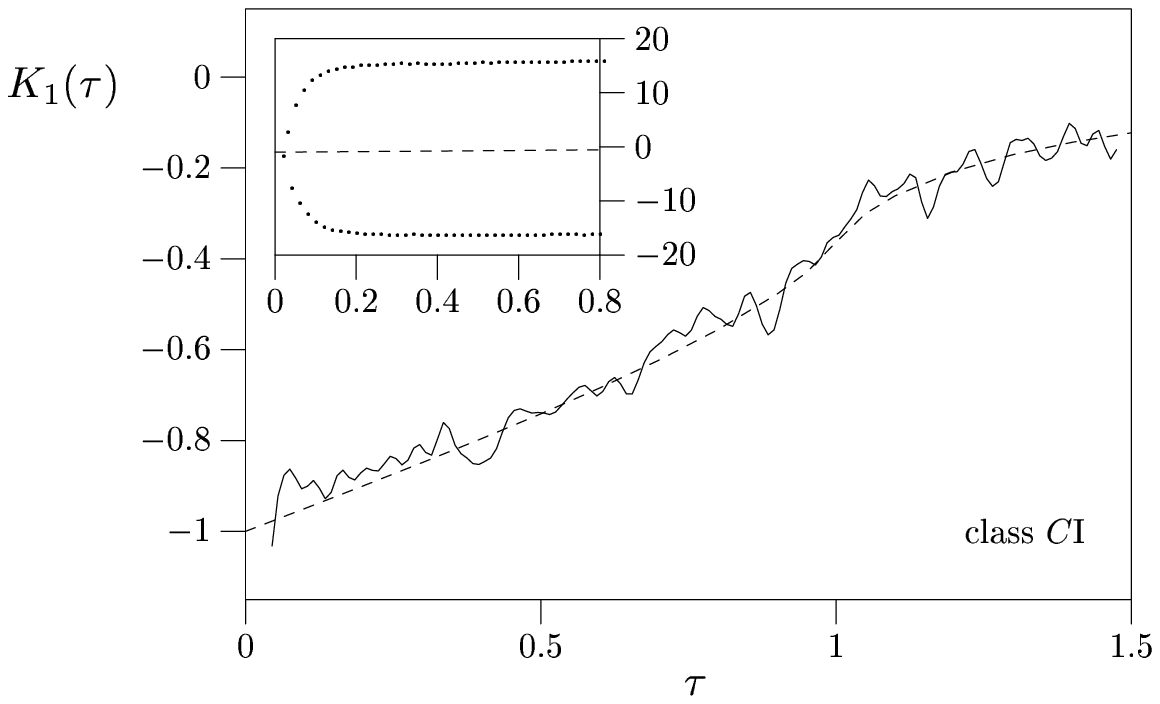}
    \caption{\label{fig:numAndreevC}  First-order form factor
      for Andreev star graphs in the classes $C$ and $C$I
      averaged over
      $10 000$ realizations with $B=100$ bonds
      (additional time average over an interval
      of length $8 \frac{g}{\mu B}$).  
      Dashed lines: prediction by Gaussian random-matrix theory. 
      Full lines: numerically calculated form factor for graphs.
      Inlets: discrete time form factor $\frac{K_{1,n}}{2}$
      as function of
      $\tau=n \frac{g}{\mu B}$ for even $n$
      ($K_{1,n}$ vanishes by construction for odd $n$).
    } 
  \end{center}
\end{figure}

Let us start with the Andreev star graphs for the classes $C$ and $C$I
where the wave function can be chosen in the
simplest case to have two components.  The
first will be called ``electron'' and the second ``hole''.  The
transfer matrix $\mathcal{L}(k)$ and the central scattering matrix
defined by
\begin{equation}
  \left.
    \begin{array}{l}
      \text{$C$}\\
      \text{$C$I}
    \end{array}
  \right\}
  :\qquad
  \mathcal{S}_C=
  \begin{pmatrix}
    \mathcal{S}_{\mathrm{DFT}} &0 \\
    0 & \mathcal{S}_{\mathrm{DFT}}^*
  \end{pmatrix}
\end{equation}
obey the symmetry condition \eqref{eq:canon_S_C}.

\begin{figure}
  \begin{center}
    \includegraphics[width=0.8\linewidth]{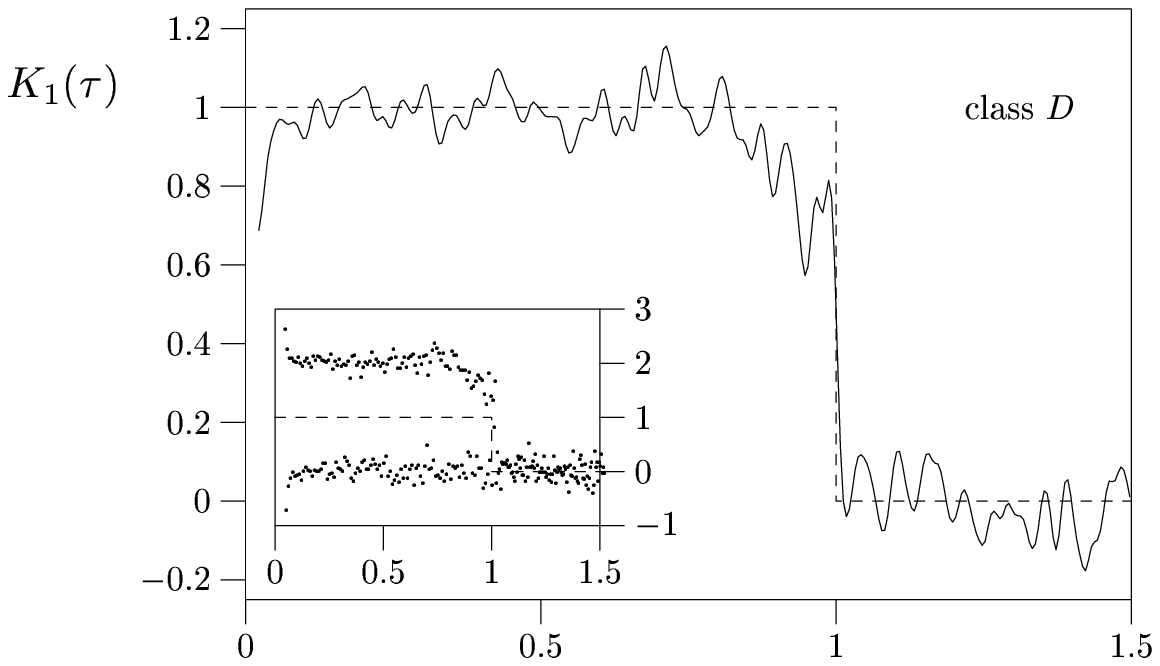}
    \includegraphics[width=0.8\linewidth]{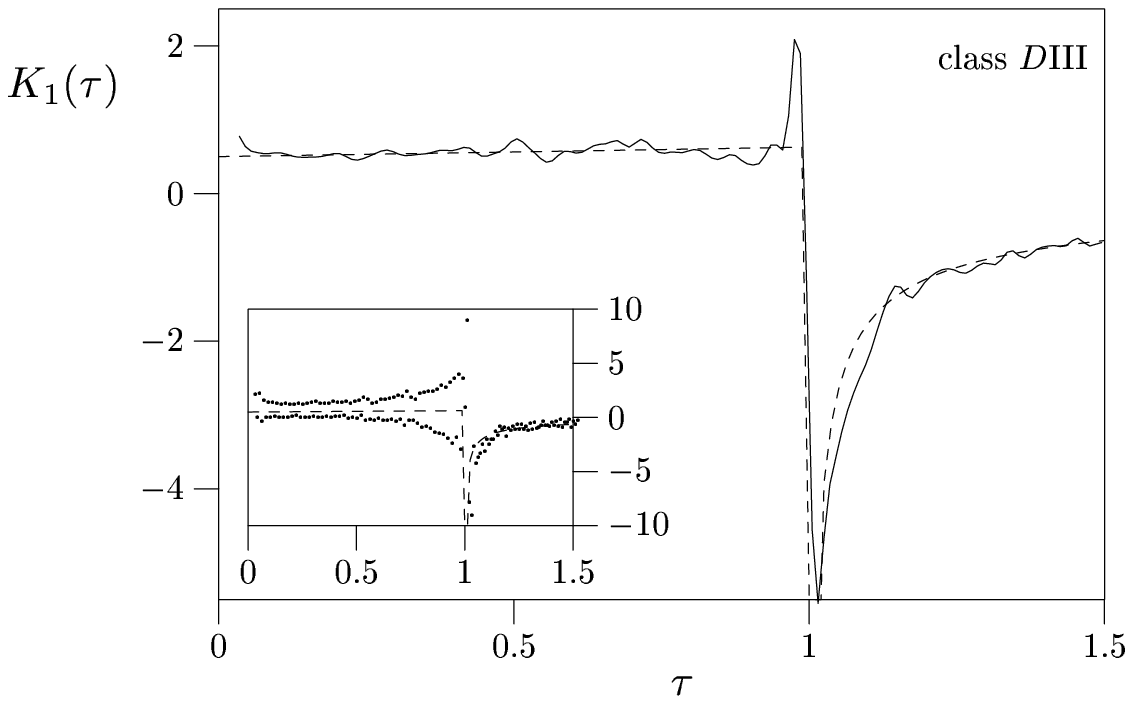}
    \caption{\label{fig:numAndreevD} 
      The first-order form factor
      for Andreev star graphs in the
      classes $D$ and $D$III (see figure
      \ref{fig:numAndreevC} for details).}
  \end{center}
\end{figure}

The
peripheral scattering matrix may be chosen such that complete 
\emph{Andreev scattering} (electron-hole conversion) 
takes place
\begin{equation}
  \left.
    \begin{array}{l}
      \text{$C$}\\
      \text{$C$I}
    \end{array}
  \right\}
  :\qquad
  \mathcal{S}_P=\frac{1}{\sqrt{2}}
  \begin{pmatrix}
    0 &\mathcal{D} \\
    -\mathcal{D}^* &0
  \end{pmatrix}
\end{equation}
where the diagonal matrix $\mathcal{D}$ is
\begin{equation}
  \text{$C$}:\qquad \mathcal{D}_{kl}=\delta_{kl} \,\Exp^{\ImgI \beta_k}
\end{equation}
for class $C$, and
\begin{equation}
  \text{$C$I}:\qquad \mathcal{D}_{kl}=\delta_{kl} \,\ImgI\sigma_k
\end{equation}
for class $C$I. The random phases $\beta_k$ are uniformly distributed
and $\sigma_k=\pm 1$ with equal probability.

For the Andreev classes $D$ and $D$III 
and as well for the three chiral classes $A$III, $BD$I, and $C$II
a four-component wave function is
needed. We will call the first (last) two
components ``electron'' (``hole'').

\begin{figure}
  \begin{center}
    \!\includegraphics[width=0.8\linewidth]{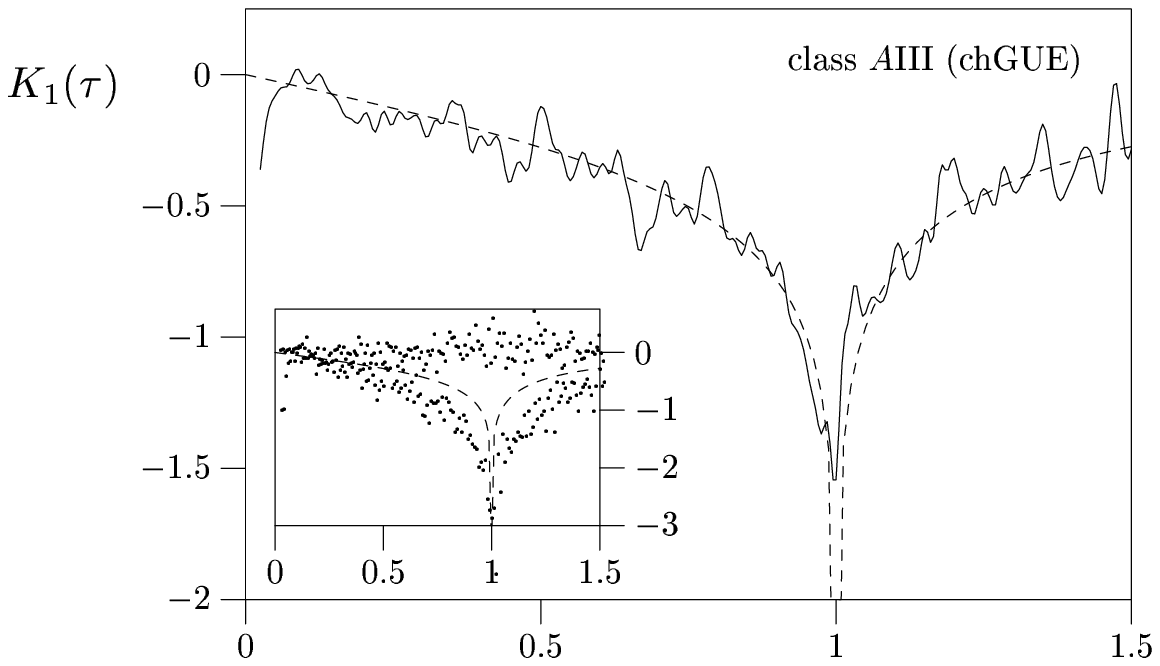}\\
    \! \includegraphics[width=0.8\linewidth]{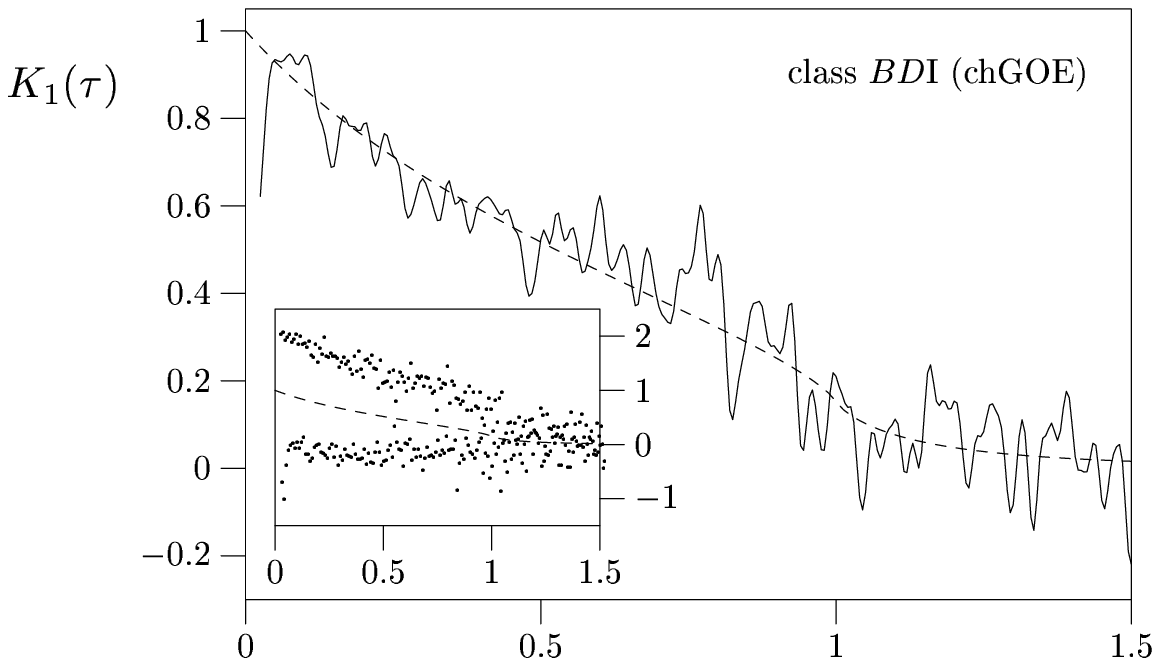}\\
    \;\;\;\includegraphics[width=0.8\linewidth]{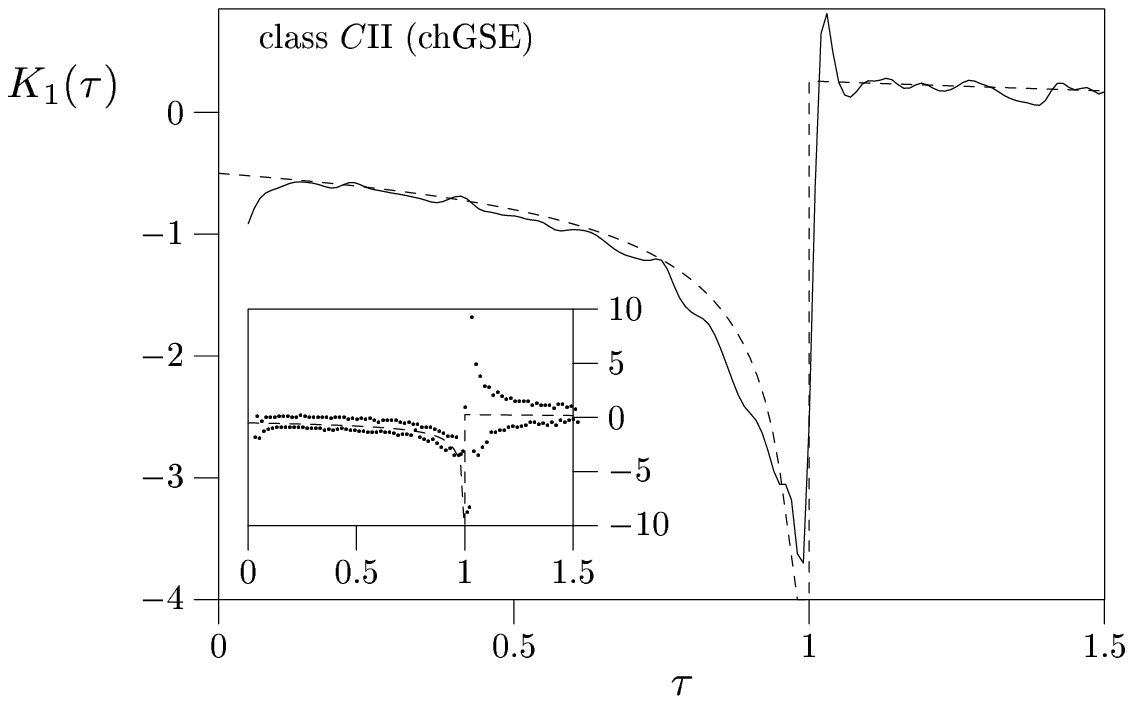}
    \caption{\label{fig:numchiral} 
      The first-order form factor
      for chiral star graphs 
      (see figure
      \ref{fig:numAndreevC} for details).
      }
  \end{center}
\end{figure}

The
symmetry requirements \eqref{eq:canon_S_D}, \eqref{eq:canon_S_DIII},
\eqref{eq:canon_S_AIII}, \eqref{eq:canon_S_BDI}, and 
\eqref{eq:canon_S_CII} are met by the transfer matrix $\mathcal{L}(k)$ 
and the central scattering matrix defined by
\begin{equation}
  \left.
    \begin{array}{l}
      \scriptstyle \text{$D$}\\
      \scriptstyle \text{$D$III}\\
      \scriptstyle \text{$A$III}\\
      \scriptstyle \text{$BD$I}\\
      \scriptstyle \text{$C$II}
    \end{array}
  \right\}
  :\;\;
  \mathcal{S}_C=
  \begin{pmatrix}
    \mathcal{S}_{\mathrm{DFT}} &0 &0&0\\
    0&\mathcal{S}_{\mathrm{DFT}} &0 &0\\
    0&0 & \mathcal{S}_{\mathrm{DFT}}^*&0\\
    0 &0&0 & \mathcal{S}_{\mathrm{DFT}}^*
  \end{pmatrix}.
\end{equation}
In all five remaining classes
we choose the peripheral scattering matrix such that complete
Andreev scattering takes place. For $D$ and $D$III the
simplest choice obeying the symmetry requirements are 
\begin{equation}
  \left.
    \begin{array}{l}
      \text{$D$}\\
      \text{$D$III}
    \end{array}
  \right\}
  :\;\;
  \mathcal{S}_P=\frac{1}{\sqrt{2}}
  \begin{pmatrix}
    0&0 &\mathcal{D}_1&\mathcal{D}_2   \\
    0&0 &-\mathcal{D}_2^*&\mathcal{D}_1^*   \\
    \mathcal{D}_1^* &\mathcal{D}_2^*&0&0\\
    -\mathcal{D}_2 &\mathcal{D}_1&0&0\\
  \end{pmatrix}
\end{equation}
where the diagonal matrices $\mathcal{D}_j$ are
\begin{equation}
  \text{$D$}:\qquad 
  \begin{array}{rl}
    \mathcal{D}_{1,kl}&=\delta_{kl} \,\Exp^{\ImgI \beta_k}\\
    \mathcal{D}_{2,kl}&=\delta_{kl} \,\Exp^{\ImgI \gamma_k}
  \end{array}
\end{equation}
for class $D$, and
\begin{equation}
  \text{$D$III}:\qquad 
  \begin{array}{rl}
    \mathcal{D}_{1,kl}&=\delta_{kl} \,\Exp^{\ImgI \beta_k}\\
    \mathcal{D}_{2,kl}&=\delta_{kl} \,\ImgI \sigma_k
  \end{array}
\end{equation}
for class $D$III. The random phases $\beta_k$ and $\gamma_k$ are
uniformly distributed and $\sigma_k=\pm 1$ with equal probability.

The simplest choice for peripheral scattering matrices 
in the chiral classes is
\begin{equation}
  \left.
    \begin{array}{l}
      \text{$A$III}\\
      \text{$BD$I}\\
      \text{$C$II}
    \end{array}
  \right\}
  :\;
  \mathcal{S}_P=\frac{1}{\sqrt{2}}
  \begin{pmatrix}
    0&0 &\mathcal{D}_1&\mathcal{D}_2\\
    0 &0&\mathcal{D}_3&-\mathcal{D}_1\\
    \mathcal{D}_4&\mathcal{D}_5&0&0\\
    \mathcal{D}_6&-\mathcal{D}_4&0&0
  \end{pmatrix}.
\end{equation}
The diagonal matrices
$\mathcal{D}_j$ have to be chosen according to the requirements of
each symmetry class. For class $A$III they are
\begin{equation} 
  \text{$A$III}:\qquad
  \begin{array}{rl}
    \mathcal{D}_{1,kl} &= \delta_{kl} \sigma_k\\
    \mathcal{D}_{2,kl} &= \delta_{kl} \Exp^{\ImgI \beta_k }\\
    \mathcal{D}_{3,kl} &= \delta_{kl} \Exp^{-\ImgI\beta_k}\\
    \mathcal{D}_{4,kl} &= \delta_{kl} \tau_k\\
    \mathcal{D}_{5,kl} &= \delta_{kl} \Exp^{\ImgI \gamma_k }\\
    \mathcal{D}_{6,kl} &= \delta_{kl} \Exp^{-\ImgI\gamma_k}
  \end{array}
\end{equation}
where $\tau_k,\sigma_k=\pm 1$ with equal probability and the phases
$\beta_k$ and $\gamma_k$ are uniformly distributed.  The $BD$I star
graphs can be obtained from the $A$III case by the additional
restrictions
\begin{equation}
  \text{$BD$I}:\qquad
  \tau_k=\sigma_k \qquad \text{and} \qquad \gamma_k=-\beta_k.
\end{equation}
Finally, for class $C$II the peripheral scattering matrix is defined
by
\begin{equation} 
  \text{$C$II}:\qquad
  \begin{array}{rl}
    \mathcal{D}_{1,kl} &= \delta_{kl} \sigma_k\\
    \mathcal{D}_{2,kl} &= \delta_{kl} \Exp^{\ImgI \beta_k }\\
    \mathcal{D}_{3,kl} &= \delta_{kl} \Exp^{-\ImgI\beta_k}\\
    \mathcal{D}_{4,kl} &= -\delta_{kl} \sigma_k\\
    \mathcal{D}_{5,kl} &= -\delta_{kl} \Exp^{\ImgI\beta_k}\\
    \mathcal{D}_{6,kl} &= -\delta_{kl} \Exp^{-\ImgI\beta_k}
  \end{array}
\end{equation}

We have checked numerically that the first-order form factor
for the constructed Andreev and chiral star graphs 
obeys the corresponding prediction
by Gaussian random-matrix theory (see figures
\ref{fig:numAndreevC},\ref{fig:numAndreevD}, and  \ref{fig:numchiral}).

\acknowledgments

We are indebted to Felix von Oppen and Martin Zirnbauer
for many helpful suggestions, comments and discussions. 
We thank for the support
of the Sonderforschungsbereisch/Transregio 12 of the Deutsche 
Forschungsgemeinschaft.

\end{document}